\documentclass[acmtog, nonacm]{acmart}
\usepackage{multirow}
\usepackage{amsmath}
\usepackage{colortbl}
\usepackage{bbding}

\newcommand{\name}{\textit{TransGS }}
\newcommand{\Gname}{\textit{GauFace }}

\begin{document}

\title{Instant Facial Gaussians Translator for Relightable and Interactable Facial Rendering}

\author{Dafei Qin}
\email{qindafei@connect.hku.hk}
\affiliation{%
  \institution{The University of Hong Kong and Deemos Technology Co., Ltd.}
  \city{Shanghai}
  \country{China}
}

\author{Hongyang Lin}
\affiliation{%
 \institution{ShanghaiTech University and Deemos Technology Co., Ltd.}
 \city{Shanghai}
 \country{China}}
\email{linhy@shanghaitech.edu.cn}

\author{Qixuan Zhang}
\affiliation{%
 \institution{ShanghaiTech University and Deemos Technology Co., Ltd.}
 \city{Shanghai}
 \country{China}}
\email{zhangqx1@shanghaitech.edu.cn}

\author{Kaichun Qiao}
\affiliation{%
 \institution{ShanghaiTech University and Deemos Technology Co., Ltd.}
 \city{Shanghai}
 \country{China}}
\email{qiaokch2022@shanghaitech.edu.cn}

\author{Longwen Zhang}
\orcid{0000-0001-8508-3359}
\affiliation{%
 \institution{ShanghaiTech University and Deemos Technology Co., Ltd.}
 \city{Shanghai}
 \country{China}}
\email{zhanglw2@shanghaitech.edu.cn}

\author{Zijun Zhao}
\affiliation{%
 \institution{ShanghaiTech University and Deemos Technology Co., Ltd.}
 \city{Shanghai}
 \country{China}}
\email{zhaozj2022@shanghaitech.edu.cn}

\author{Jun Saito}
\email{jsaito@adobe.com}
\affiliation{%
  \institution{Adobe Research}
  \city{Seattle}
  \country{USA}
}

\author{Jingyi Yu}
\affiliation{%
 \institution{ShanghaiTech University}
 \city{Shanghai}
 \country{China}}
\email{yujingyi@shanghaitech.edu.cn}

\author{Lan Xu}
\affiliation{%
 \institution{ShanghaiTech University}
 \city{Shanghai}
 \country{China}}
\email{xulan1@shanghaitech.edu.cn}

\author{Taku Komura}
\authornote{Corresponding author}
\email{taku@cs.hku.hk}
\affiliation{%
  \institution{The University of Hong Kong}
  \city{Hong Kong}
  \country{China}
}

\begin{abstract}
The advent of digital twins and mixed reality devices has increased the demand for high-quality and efficient 3D rendering, especially for facial avatars. Traditional and AI-driven modeling techniques enable high-fidelity 3D asset generation from scans, videos, or text prompts. However, editing and rendering these assets often involves a trade-off between offline quality and online speed. In this paper, we propose \Gname, a novel Gaussian Splatting representation, tailored for efficient animation and rendering of physically-based facial assets. Leveraging strong geometric priors and constrained optimization, \Gname ensures a neat and structured Gaussian representation suitable for efficient rendering and generative modeling. \\
Then, we introduce \name, a diffusion transformer that instantly translates physically-based facial assets into the corresponding \Gname representations. Specifically, we adopt a patch-based pipeline to handle the vast number of Gaussians effectively. We also introduce a novel pixel-aligned sampling scheme with UV positional encoding to ensure the throughput and rendering quality of \Gname assets generated by our \name. Once trained, \name can instantly translate facial assets with lighting conditions to \Gname representation, delivering high fidelity and real-time facial interaction of 30fps@1440p on a \textit{Snapdragon\textsuperscript{\textregistered} 8 Gen 2} mobile platform. Notably, with the rich conditioning modalities, it also enables editing and animation capabilities reminiscent of traditional CG pipelines. \\
We conduct extensive evaluations and user studies, compared to traditional offline and online renderers, as well as recent neural rendering methods, which demonstrate the superior performance of our approach for facial asset rendering. We also showcase diverse immersive applications of facial assets using our \name approach and \Gname representation, across various platforms like PCs, phones and even VR headsets.

\end{abstract}

\begin{teaserfigure}
  \includegraphics[width=\textwidth]{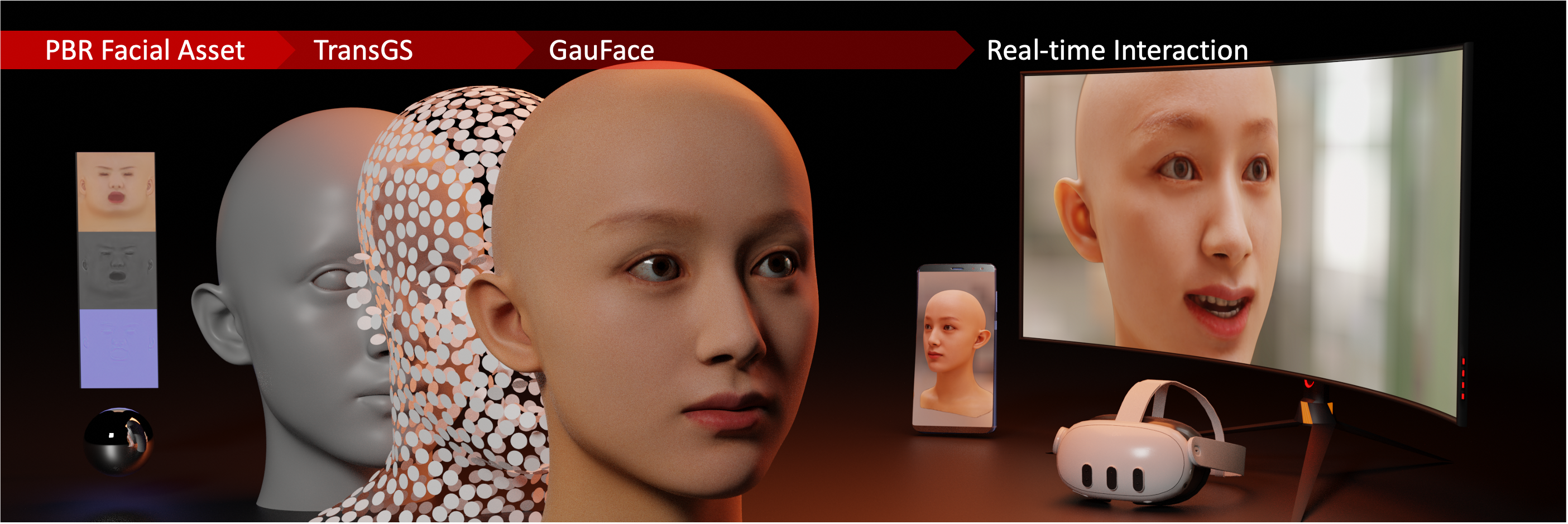}
  \caption{Conditioning on a desired lighting, our proposed \name system transfers physically-based facial assets (left) to \Gname (middle), a novel Gaussian representation, in \textit{seconds}. The resulting \Gname asset offers high-quality real-time rendering and animation across various platforms (right). }
  \label{fig:teaser}
\end{teaserfigure}

\maketitle

\section{Introduction}
\label{sec:intro}

The advent of digital twins and mixed reality (MR) devices is transforming expectations for 3D asset quality and rendering efficiency. Given their central role in human interaction, facial avatars, which convey emotions and intentions, must be rendered with precision. Our innate ability to perceive even the slightest inaccuracies in facial animations can lead to the uncanny valley effect, where almost but not entirely lifelike appearances cause viewer discomfort.

Production-level workflows for crafting lifelike facial avatars involve both modeling and rendering. The modeling phase aims to recover CG-friendly facial assets with nuanced facial idiosyncrasies. These include realistic geometry with structured UV unwrapping, physically-based appearance (e.g., diffuse albedo, specular intensity, roughness, normal map, and displacement), and the motion rig to empower expression manipulation and animation. The past decade has witnessed the rapid progress of such facial modeling. Early attempts~\cite{DigitalEmily, debevec2000acquiring} require immense artistic sculpting or expensive apparatus like Light Stage. Recent advances in AI-Generated-Content lower the burdern of facial modeling. One can easily customize physically-based facial assets from more light-weight inputs, i.e., a single scan~\cite{li2020dynamic}, a monocular video or image~\cite{cao2022authentic}, or even text prompts~\cite{zhang2023dreamface}.
Yet, the rendering phase of such facial assets has been left behind. Thus far, most rendering tools in existing CG software adopt the GPU-based rasterization pipeline, offline or online. Offline renderers like Arnold~\cite{arnold2024} or Cycles~\cite{blender2018}, are characterized by intricate ray tracing computations, frequently leading to prolonged rendering duration. Conversely, online ones like Unity3D Build-in Pipeline or OpenGL prioritize interactive responsiveness at the expense of rendering fidelity. Hence, this leads to the substantial distinction between facial avatars featured in offline cinematic productions and online gaming environments. Take the distinct characters of Keanu in the feature film "The Matrix" and the game "Cyberpunk 2077" as examples.

Recent neural advances~\cite{mildenhall2021nerf,NR-survey} conduct volume rendering at photo-realism, bringing new potentials to narrow the gap between traditional offline and online rendering. Notably, 3D Gaussian Splatting (3DGS)~\cite{kerbl20233d} stands out for its exceptional rendering ability at speed and quality. Also, its explicit representation can be seamlessly integrated into the GPU-based rasterization pipeline. Various approaches extend 3DGS into dynamic scenes, i.e., human performance~\cite{hifi4G,4dgs} or facial animations~\cite{saito2023relightable, qian2023gaussianavatars, chen2023monogaussianavatar}. 
Yet, they rely on per-scene training for modeling real-life scenes from video or image inputs, thus unable to directly apply to CG-friendly facial assets for instant improvement of rendering quality. 
One can apply offline render techniques, e.g. monte carlo path-tracing and subsurface-scattering on the facial assets to create high-quality images under diverse expressions and view angles and subsequently optimize the 3DGS for render acceleration. But such cumbersome data preparation and optimization make it impractical for interactive applications, let alone supporting lighting setup, online editing, or animation control as if using the original facial assets.

To tackle the above challenges, we propose \name, a novel translator to instantly translate CG-friendly facial assets into novel 3DGS-like representations. As shown in Fig.~\ref{fig:teaser}, our approach enables real-time and high-quality rendering of physically-based facial assets, even comparable with offline rendering techniques. It also can be seamlessly integrated into various platforms like PC, phone, or VR headset, and it is compatible with traditional facial assets for relighting, editing, and animation capabilities. 

The key design in our approach is a Gaussian representation tailored for facial assets, dubbed \Gname. On the one hand, our \Gname bridges Gaussian splatting with the strong geometry and texture priors in the facial assets. On the other hand, careful optimization yields a neat and structured representation, paving the way for adopting an efficient generative paradigm in our translator.
Specifically, in our \Gname, we attach and rig the Gaussian primitives onto the face mesh to naturally support the mesh-based animation of the original facial asset. To model nuanced appearance changes under animation, we adopt a novel shading vector to disentangle the deformation-dependent/agnostic shading effects.   
Then, for a facial asset with a desired lighting map, we tailor the optimization scheme from 3DGS to obtain the corresponding \Gname asset. We adopt a \textit{Pixel Aligned Sampling} scheme to uniformly initialize Gaussians on the UV plane and defer the pruning operation during optimization until rendering. Such strategies not only constrain Gaussian attributes to have neat and tight distributions but also make their sampling positions known to provide rich conditioning priors for our generative translator.

Then, we develop our instant generative translator \name with a diffusion transformer (DiT) architecture to model the intricate one-to-many relationship between a 3D facial asset and its \Gname transcript. To train \name, we first collect 143 physically-based facial assets with 4K textures and geometry and render them under 134 HDR environment maps, resulting in a total of 1,023 combinations. We prepare 1,071 high-quality rendered images for each combination to optimize the corresponding \Gname asset. We then train \name on the optimized 1,023 \Gname assets. During training, we adopt a patch-based pipeline of DiT architecture to effectively handle the vast number of Gaussian points per \Gname asset. We also utilize UV positional encoding to guide the transformer's attention toward textures proximal to the Gaussian sampling positions. With rich conditioning inputs on the geometry, PBR textures of the facial asset, and the desired HDR environment lighting, we can swiftly generate the \Gname transcript of a facial asset in a matter of seconds. 
Thanks to such efficient generation and rich conditions, \name supports swiftly transferring new modifications of geometry and image textures from the original facial assets to the generated \Gname assets, thereby offering control and editing capabilities akin to traditional CG pipelines. We conduct extensive evaluations, qualitative and quantitative, to validate the effectiveness and the rendering quality of our approach. We demonstrate superior rendering results to traditional online rendering in common CG software, even close to the quality of offline ones. We also showcase the capability of \name on a series of immersive applications on various platforms, i.e., multi-source conditioned generation, cross-platform interactions, and rapid editing.

\section{Related Works}
\label{sec:related}

\subsection{Face Rendering}
\paragraph{Traditional Rendering Techniques}

A shading model is necessary for rendering photorealistic images. Phong shading\cite{phong1998illumination} interpolates surface normals for smooth highlights, followed by Physically Based Rendering (PBR)\cite{pharr2016physically} which models realistic light interactions, and later, Bidirectional Surface Scattering Reflectance Distribution Function (BSSRDF)\cite{jensen2001practical} extends this by considering light that penetrates and scatters within surfaces. Some following works\cite{hanrahan2023reflection, borshukov2005realistic, habel2013photon} focus on facial rendering, which significantly influence the color and realism of facial assets in digital imagery.  \citet{dEon2007RealTimeSkin} enhances the realism of specular reflections on the skin, adapting them to different lighting conditions and angles. 
\citet{donner2008layered} introduces inter-layer absorption to get higher skin rendering accuracy. \citet{pixar2016sss} introduces exponential and non-exponential path tracing to simulate more accurate photon behavior within human skin, enhancing realism and detail in rendered images.

\paragraph{Volume Rendering}
Several studies have explored the application of Neural Radiance Fields (NeRF)~\cite{mildenhall2021nerf} for facial reconstruction and rendering.    \citet{gafni2021dynamic} utilizes deformable 3D Morphable Models (3DMMs)~\cite{blanz2023morphable} to capture dynamic facial movements.~\citet{athar2022rignerf, gao2022reconstructing} enhance the modeling capabilities by incorporating additional Multilayer Perceptrons (MLPs) to account for deviations from the 3DMM-defined space.~\citet{grassal2022neural, zheng2022avatar} introduce view and expression-dependent textures to achieve detailed textural fidelity. Furthermore,~\citet{lombardi2021mixture} boosts rendering speeds by substituting traditional NeRFs with volumetric primitives, while~\citet{ma2021pixel} utilizes a rendering-adaptive per-pixel decoder for enhanced rendering efficiency and~\citet{zielonka2023instant} significantly reduces optimization times by leveraging Instant-NGP~\cite{muller2022instant}.  
However, while these techniques excel in reconstructing real-life scenes, their applicability to the translation of CG assets is constrained. Besides, as an implicit representation, NeRF and its variants are not directly compatible with existing graphics pipelines, preventing their board applications.

\begin{figure*}[htb]
    \centering
    \includegraphics[width=\linewidth]{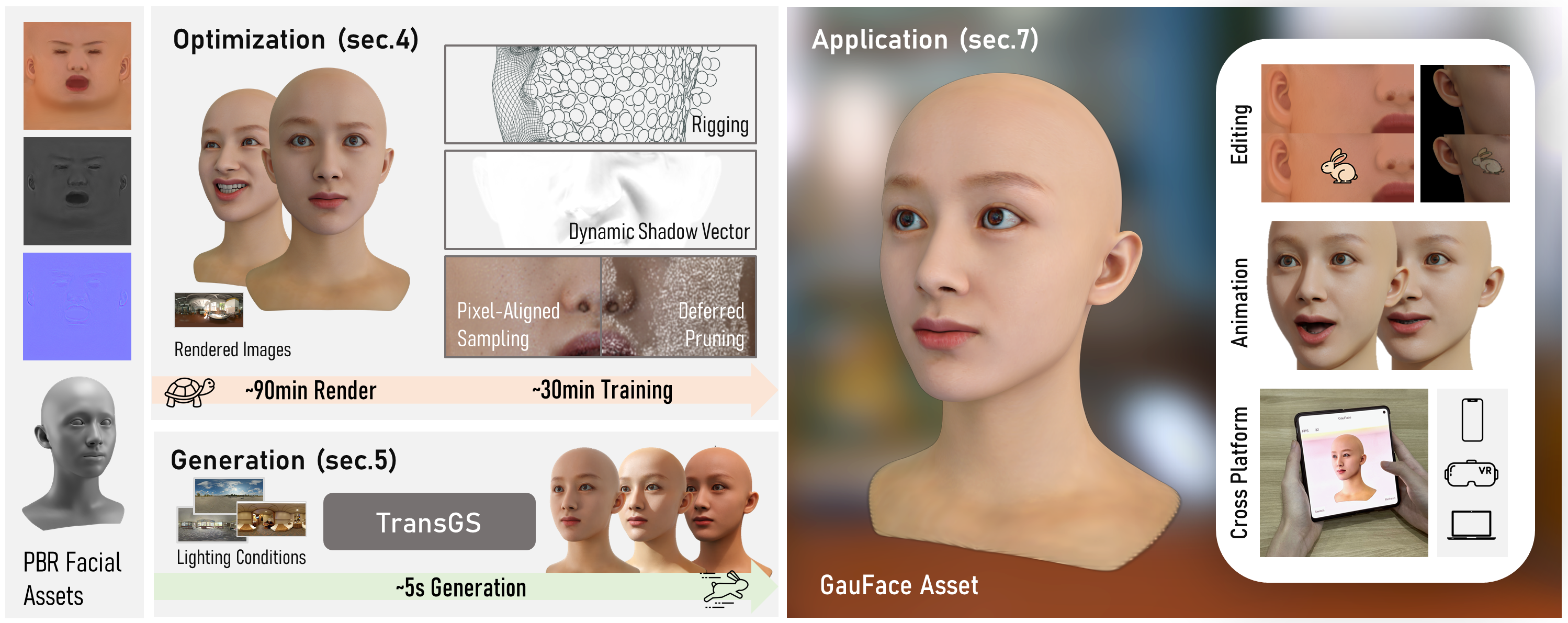}
    \caption{\textbf{Overview.} We present two methods for obtaining relightable dynamic Gaussian facial assets. The first method (Sec. \ref{sec:gs_face}) render high-quality multi-view images and optimize the \Gname representation. The second method (Sec. \ref{sec:gs_gen}), which we introduce as \name, directly generates \Gname assets from textures and models in approximately 5 seconds. } 
    \label{fig:overview}
\end{figure*}

\paragraph{3DGS Variants} Recently, 3D Gaussian Splatting (3DGS)~\cite{kerbl20233d} proposed an explicit volume rendering pipeline, achieving both high-quality rendering and real-time performance.  Research has explored various techniques to enhance 3DGS in avatar reconstruction and animation. Serval works ~\cite{shao2024splattingavatar, qian2023gaussianavatars, ma20243d, rivero2024rig3dgs} use geometry-based parameterization to rig 3D Gaussians. 
Implicit representations like neural networks are introduced~\cite{xiang2024flashavatar, xu2023gaussian, saito2023relightable, dhamo2023headgas} to enlarge the representation capacity of 3DGS. SplatFace~\cite{luo2024splatface} introduces a constrained splat-to-surface method for better facial animation.  PSAvatar~\cite{zhao2024psavatar} uses points-based geometry representation to fit the face deformation better. GaussianHead~\cite{wang2023gaussianhead} embeds the Gaussian features to a multi-scale tri-plane structure to increase the representation power. However, these methods either disregard deformation-dependent color information~\cite{qian2023gaussianavatars, shao2024splattingavatar} or incorporate additional neural networks for fitting~\cite{dhamo2023headgas, saito2023relightable, xu2023gaussian}, posing challenges in balancing rendering quality and efficiency.

\subsection{Face Generation}

\paragraph{PBR Facial Asset Generation} The development of 3D face generation was notably advanced by AvatarMe~\cite{lattas2020avatarme} and other improved versions~\cite{luo2021normalized, lattas2023fitme}, which offer photorealistic 3D avatars that integrate well with conventional computer graphics pipelines.~\citet{li2020dynamic} introduces an end-to-end framework to automate the generation of high-quality facial assets and rigs using mesh and UV maps.~\citet{cao2022authentic} employs a universal avatar prior, trained on a vast dataset, to facilitate high-quality facial asset creation from simple phone scans.~\citet{zhang2023dreamface} generate PBR facial assets conditioning on images or text prompts. 

\paragraph{GAN / NeRF Based Generation.}
Several works use GAN~\cite{isola2017image, karras2019style, hou2021towards, hou2022face, papantoniou2023relightify,  karras2020analyzing} and diffusion-based methods~\cite{ponglertnapakorn2023difareli, zhang2024facednerf} to generate different views and relightable face images. To improve 3D consistency, NeRF has been widely adopted~\cite{schwarz2020graf, chan2021pi, gu2021stylenerf}, with parametric human head models~\cite{hong2022headnerf, zhuang2022mofanerf} to disentangle the rendering pose, identity, expression, and appearance. Implicit geometry structures like Signed Distance Function (SDF)~\cite{or2022stylesdf} and tri-plane~\cite{chan2022efficient} are leveraged for better 3D consistency.  Subsequently,~\citet{wang2023rodin} employs a diffusion model to generate tri-plane-based 3D models. While the transition from GAN-based image generators to tri-plane-based models has substantially increased 3D coherence, it still falls short of explicit 3D representations like meshes and voxel grids.

\paragraph{GS-Based Generation}
3DGS has been adapted to generate general 3D models~\cite{liu2023humangaussian, tang2023dreamgaussian, yi2023gaussiandreamer}. For instance, DreamGaussian combines a generative 3DGS model with mesh extraction and texture refinement for better visual quality. \citet{zhou2024headstudio} introduce HeadStudio, which innovatively creates animatable avatars from textual prompts by combining 3DGS with FLAME-based 3D representations.  However, the rendering images of HeadStudio-generated assets are overly saturated, and the animation quality suffers due to the reliance on the SDS loss.

\subsection{Post-Editing}
\paragraph{NeRF editing} Recent work in 3D modeling addresses the editing limitations of NeRF-based and GAN-based methods, enhancing customization capabilities. For instance, FENeRF~\cite{sun2022fenerf} focuses on generating consistent and editable 3D portraits, IDE-3D~\cite{sun2022ide} enables precise adjustments to individual facial attributes.~\citet{aneja2023clipface} uses the GAN model and~\citet{yue2023chatface} uses diffuse-based models to modify facial images by text prompts SketchFaceNeRF~\cite{lin2023sketchfacenerf} offers an intuitive interface for sketch controlled face generation. However, in these methods, all the modifications are done in the neural latent space, resulting in imprecise control and resolution bottleneck. Besides, these methods are not compatible with the CG industry, where modifications are performed by editing the explicit geometry, image textures, etc.

\paragraph{3DGS Editing} Recent 3DGS-based editing methods like~\citet{fang2023gaussianeditor} and~\citet{chen2023gaussianeditor} introduce innovative text-based techniques for editing 3D Gaussian representations, facilitating precise adjustments to shapes and textures. However, to perform the editing, these methods require additional Gaussian optimizations, which is cumbersome for timely preview and iterations.

\section{Preliminary}
\label{sec:prelim}
\subsection{Physically-Based Rendering Facial Assets}
A Physically-Based Rendering (PBR) facial asset is a digital representation of a human face optimized for rendering using a physically accurate lighting and shading process. 
In this paper, we define a PBR facial asset as a collection of mesh neural geometry $G$, image textures $I$, including the diffuse map, normal and specular map, and expression blendshapes $\mathcal{B}$. 

We define the UV mapping function as follows: 
\begin{equation}
    (x, y, z) = M(u, v;G),
\end{equation}
which maps the points on the 2D UV space with coordinates $\boldsymbol{\mu}=(u, v)$ to the 3D position $(x, y, z)$ on the surface of neural mesh geometry $G$.

\subsection{3D Gaussian Splatting}
3DGS represents the 3D scene via a set of explicit 3D Gaussians. Each 3D Gaussian has multiple learnable parameters, including the center $\boldsymbol{\mu}$, covariance $\boldsymbol{\Sigma}$, opacity $\sigma$, and view-dependent colors represented by Spherical Harmonics $\boldsymbol{c}$. The shape of each 3D Gaussian is defined as:
\begin{equation}
    X = e^{-\frac{1}{2}\boldsymbol{\mu\Sigma}^{-1}\boldsymbol{\mu}}.
\end{equation}

When rendering a novel view, the 3D Gaussians are projected to 2D screen space ordered by depth. The color $C$ of a pixel is computed by $\alpha$-blending of these projected Gaussians:
\begin{equation}
    C=\sum_{i\in N}c_{i}\alpha_{i}\prod_{j=1}^{i-1}(1-\alpha_{j}),
\end{equation}
where $c_i$ is the evaluated view-dependent colors.

During optimization, the gradients of all attributes are computed by comparing the difference between the rendered image and the ground truth via differentiable rendering. Pruning and densification are applied according to the gradient statistics, enabling efficient scene representation and fast optimization. 

Next, as illustrated in Fig.~\ref{fig:overview}, we first introduce the design and optimization of \Gname in Sec.~\ref{sec:gs_face}. Following that, we present our generative model, \name, in Sec.~\ref{sec:gs_gen}. We then conduct a detailed evaluation in Sec.~\ref{sec:eval} and discuss the broad applications empowered by our approach in Sec.~\ref{sec:app}.

\section{\Gname: Bridging PBR Facial Assets to 3D Gaussian Splatting}
\label{sec:gs_face}
\begin{figure*}[t]
    \centering
    \includegraphics[width=\linewidth,trim={70 210 250 110},clip]{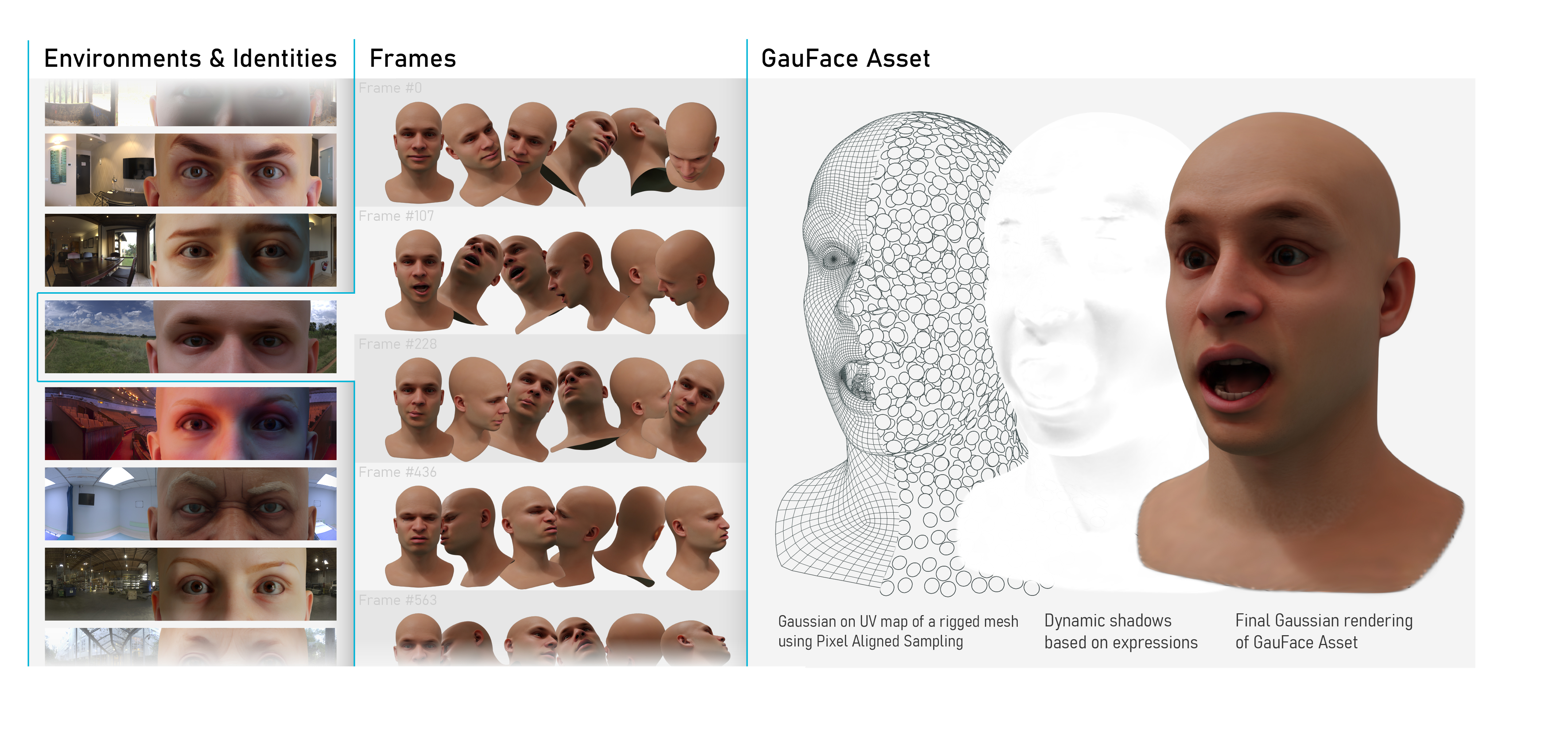}
    \caption{\textbf{PBR facial assets and \Gname representation.} \textit{Left:} We collect 143 facial assets under 134 lighting conditions, with a total of 1,023 combinations. \textit{Middle:} For each combination, we render 1,071 frames under 153 different expressions with random camera positions. \textit{Right:} Our \Gname asset defines the center of Gaussians on the UV map consistent across different identities and introduces dynamic shadow vectors to disentangle the deformation-dependent and deformation-agnostic shading effects.}
    \label{fig:gs_arch}
\end{figure*}

Recently, several works have extended 3DGS to handle facial animation~\cite{chen2023monogaussianavatar, wang2023gaussianhead, saito2023relightable}. However, they focus on reconstructing from the real world and novel view synthesis. There lacks an efficient method for converting  CG assets to 3DGS for efficient rendering.

We introduce \Gname to bridge the gap between fine-grained PBR facial assets and high-quality and efficient 3DGS representation.  Leveraging a uniform topology and UV mapping common in CG facial asset libraries~\cite{li2020learning}, as shown in Fig.\ref{fig:gs_arch}, we define Gaussian points within the shared UV space. This approach leads to a natural support of blendshape-based facial animation. To decouple the deformation-dependent shading effects from the deformation-agnostic parts, we construct \textit{Dynamic Shadow Vector} to efficiently adapting to the dynamic lighting and shadows produced by facial animation. 

Specifically, a \Gname asset is a collection of $N$ orderless Gaussian points $A = \{p_i\}_{i=1}^N$. Each Gaussian point $p$ contains the following parameters:
\begin{equation}
    p = (\boldsymbol{\mu}, d, \boldsymbol{s}, \theta, \sigma, \boldsymbol{c}, \boldsymbol{l}),
\end{equation}
where $\boldsymbol{\mu}=(u, v)$ is the Gaussian position defined on the UV space, $d$ is the deviation to the mesh surface, $\boldsymbol{s}$ and $\theta$ define the shape of the Gaussian point, $\sigma$ is the opacity of the Gaussian point, $\boldsymbol{c}$ is the color represented via Spherical Harmonics, and $\boldsymbol{l}$ is a dynamic shadow vector to represent the deformation-dependent shadings.

\subsection{Rigging 3D Gaussians}
\label{subsec:rigging}

We rig 3D Gaussians to the mesh surface to support facial animation. 
The 3D position of a Gaussian point $p$ is obtained as follows:
\begin{equation}
\label{eq:3dpos}
    \mu_{\text{3D}} = M(\boldsymbol{\mu};G) + d\boldsymbol{n}_f,
\end{equation}
where $M$ is the UV mapping function, $\boldsymbol{\mu}$ is the UV position, $f$ is the contained triangle, $d$ is a learnable parameter and $\boldsymbol{n}_f$ is the normal vector of the triangle.  This defination allows the Gaussian point to move according to the mesh deformation, and also deviate from the mesh surface along the normal direction.

We follow \citet{guedon2023sugar, huang20242d} to constrain $p$ as thin shells, by setting the scaling vector as follows: 
\begin{equation}
\label{eq:scale}
    \boldsymbol{s}=[\epsilon, s_1, s_2],
\end{equation}
where $\epsilon$ is a pre-defined small value, and $s_1, s_2$ are optimizable parameters. 
For rotation, we follow \cite{guedon2023sugar} to first calculate the rotation matrix of triangle $f$ as $R_f = [R^{(0)}, R^{(1)}, R^{(2)}]$. Then, the rotation matrix of $p$ is defined as follows:
\begin{equation}
    \label{eq:3drot}
    R_p = [R^{(0)}, xR^{(1)}+yR^{(2)}, -yR^{(1)}+xR^{(2)}]
\end{equation}
where $x+iy$ is a normalized complex number. Eq.~\ref{eq:scale} and Eq.~\ref{eq:3drot} constrain 3D Gaussians to be thin shells attaching to the mesh surface, where two degrees of freedom ($s_1, s_2$) are allowed to control the size of 3D Gaussians on the tangent space of their attached triangles, and one degree of freedom ($\theta$) is allowed to rotate along the normal direction of their attached triangles. 

Since our 3D Gaussians are attached to a mesh, it naturally supports blendshape animations by updating the mesh vertex positions at every frame. Given $B$ different blendshape vectors, the mesh vertex positions at frame $i$ are updated as follows:
\begin{equation}
    V_i = \Bar{V} +\mathcal{B}\boldsymbol{b}_i,
\end{equation}
where $\Bar{V}$ is the mesh vertices of neural geometry $G$,  $\mathcal{B}$ is the blendshape matrix and $\boldsymbol{b}_i\in[0, 1]^B$ is the the blendshape weights at frame $i$. The 3D position of each 3D Gaussian at each frame is updated via Eq.~\ref{eq:3dpos} and the rotation matrix is updated via Eq.~\ref{eq:3drot}. 

\subsection{Dynamic Shadow Vector for Deformation-Dependent Shading}
\label{subsec:shadow}
To adapt 3DGS to handle facial animation, deformation-dependent shading effects need to be supported. Recent methods either omit this effect~\cite {qian2023gaussianavatars, shao2024splattingavatar}, or use neural networks to infer frame-dependent colors~\cite{saito2023relightable, xu2023gaussian}. The latter approach entangles deformation-dependent shading effects with deformation-agnostic parts, which hinders the generalizability to unseen facial expressions. Instead, we fully leverage the geometry prior provided by the PBR facial dataset, and explicitly define attributes for the deformation-dependent shading effects.

We observe that for a face under a specific lighting condition, the change of shadows and self-occlusions contributes most to the deformation-dependent colors. These effects are mainly view-agnostic. Thus, for each Gaussian point $p$, we define a shadow vector $\boldsymbol{l}\in [0, 1]^{B}$ to express these changes caused by the facial deformations. The final color of a Gaussian point is calculated as follows:
\begin{equation}
    c^*_{\boldsymbol{\omega}, i} = \boldsymbol{l}^T \boldsymbol{b}_i~c_{\boldsymbol{\omega}},
\end{equation}
where $\boldsymbol{b}_i$ is the blendshape weight of frame $i$, and $c_{\boldsymbol{\omega}}$ is the color expressed by Spherical Harmonics at camera view $\boldsymbol{\omega}$. $\boldsymbol{l}$ is optimized in the same way as the SH color attributes $\boldsymbol{c}$, receiving gradients from differentiable rendering. 
Introducing the dynamic shadow vector for deformation-dependent shading only adds $B$ new parameters to each Gaussian point, which maintains the efficiency of the color representation. 

\subsection{Constraining \Gname for Generative Modeling}
\label{subsec:pixel}
The original 3DGS optimization algorithm initializes very few  Gaussian points in the beginning and gradually densifies them according to the complexity of the target scene. Although this design ensures high efficiency of the 3DGS representation, it makes the optimized Gaussian asset too scene-specific to form a learnable distribution.

\paragraph{Pixel Aligned Sampling} To constrain the distribution of Gaussian points, in \Gname, we sample uniformly on the UV plane during initialization, and then lock the position and turn off densification and pruning operations during optimization.  We call this \textit{Pixel Aligned Sampling}. This strategy guarantees that all different Gaussian assets share the same number of Gaussian points and their UV positions after optimization, which leads to a unified sampling pattern across all different assets. 

\begin{figure}[t]
    \centering
    \includegraphics[width=\linewidth]{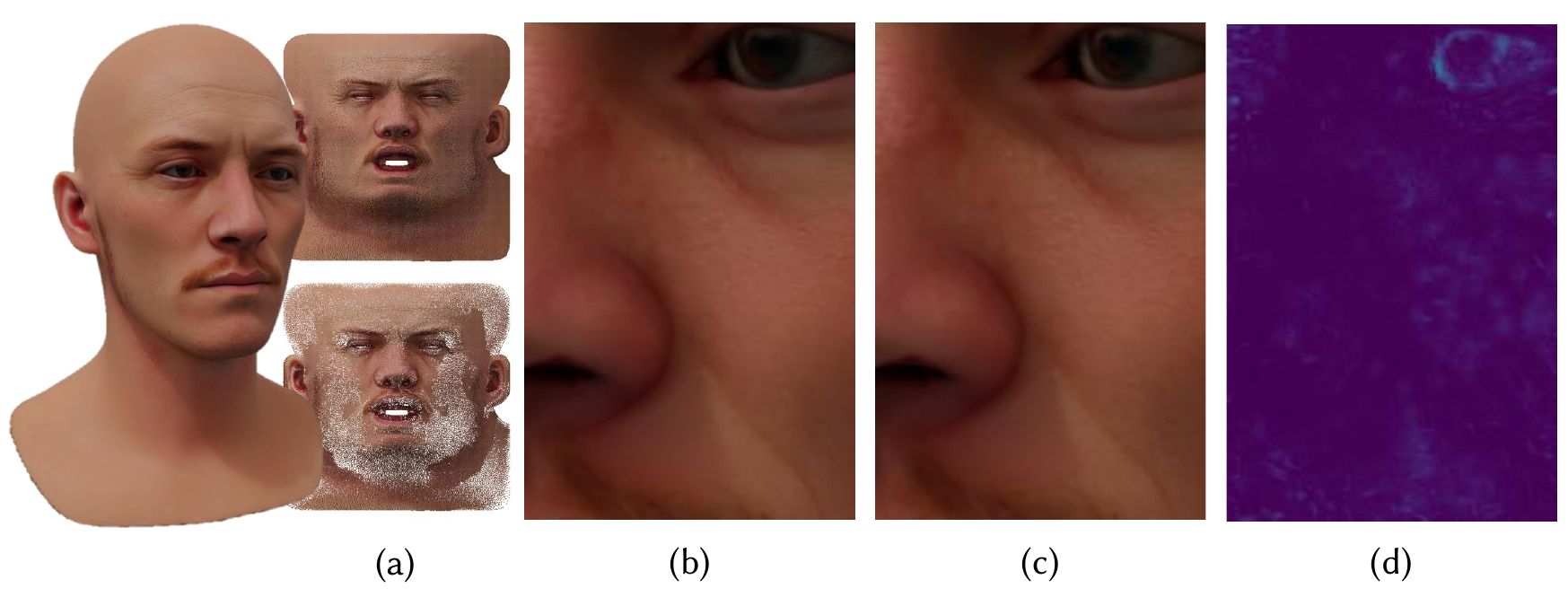}
    \includegraphics[width=0.8\linewidth]{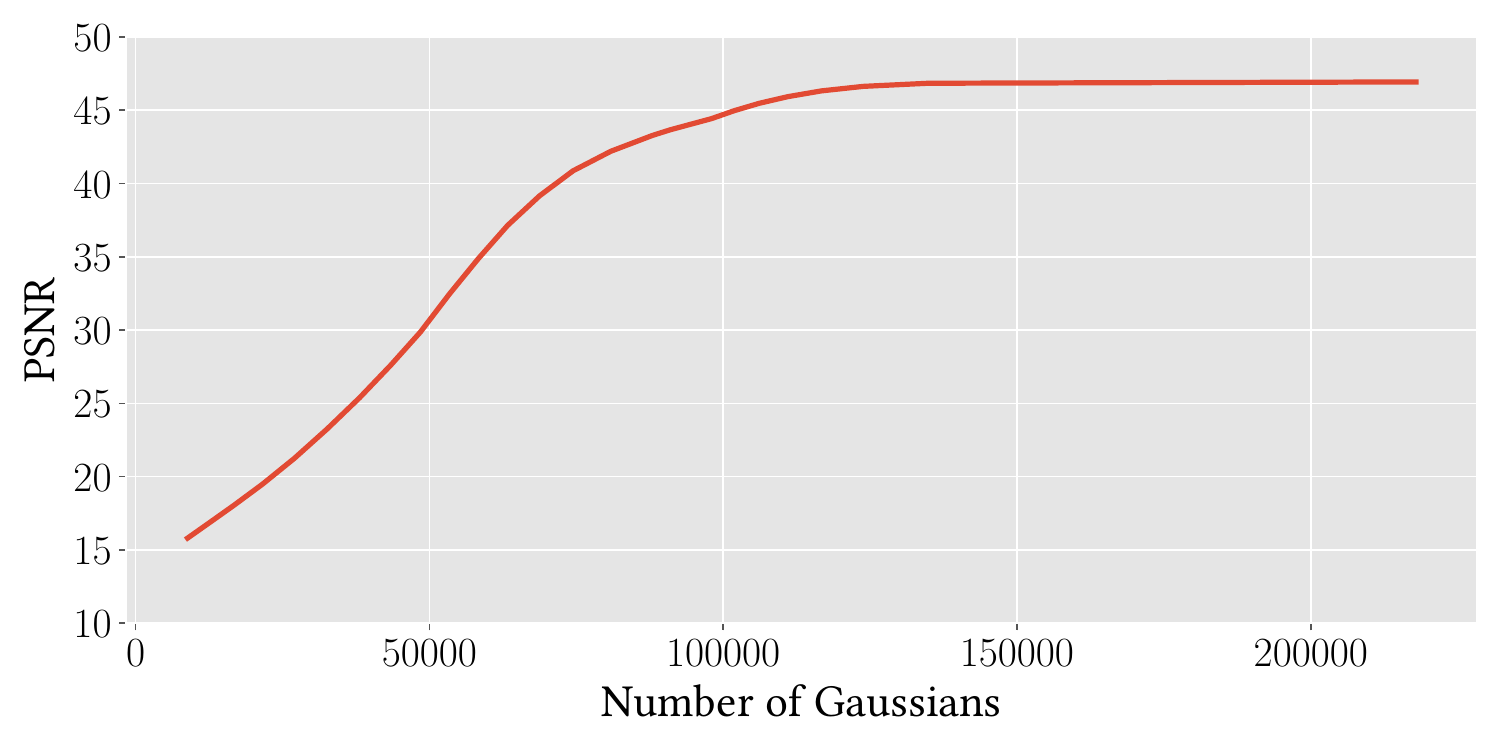}
    \caption{\textbf{Deferred Pruning.} \textit{Upper}: (a) \Gname and it's UV space base color visualization before and after pruning. (b) Without pruning, 22k points. (c) Pruned with opacity $\sigma \le 0.1$, 11k points. (d) Difference between (b) and (c). \textit{Lower}: PSNR vs. Number of Gaussians Curve. PSNR is calculated on 306 testing images with different expressions and camera positions. Gaussian points are pruned with different opacity value.}
    \label{fig:gs_curve}
\end{figure}

\paragraph{Deferred Pruning} Although \textit{Pixel Aligned Sampling} provides a rigid pattern for generative modeling, it sacrifices the rendering efficiency. Because the initialization is over-parametrized in many face parts, and pruning of unnecessary Gaussian points is disabled. However, we found that during optimization, unnecessary Gaussian points on over-parameterized areas lead to close-to-zero opacity values. This is because the opacity of all Gaussian points will be reset to zero once every several iterations, and unnecessary Gaussian points then receive little image-based gradients since the area is already represented well.  As shown in Fig.~\ref{fig:gs_curve}, after optimization finished, nearly half of the Gaussians can be pruned via opacity thresholding with minimum visual degradation. Thus, we can do pruning just before the runtime. In this way, we can keep the Gaussian point sampling positions as a prior during generative modeling while achieving similar rendering efficiency as the vanilla 3DGS.

\subsection{\Gname Dataset}
\label{subsec:gauface_dataset}
We propose the \Gname dataset to construct a mapping from PBR facial assets to their \Gname counterparts. We collect 143 PBR facial assets generated from \citet{zhang2023dreamface}, including the meshes, 51 ARKit blendshapes, and diffuse, normal, and specular maps in 4K resolution. Each asset contains 5 components: \textit{Foreface, Backhead, Teeth, Lefteye} and \textit{Righteye}, which share the same topology as ICT-FaceKit~\cite{li2020learning}.  We render the facial assets under 134 different lighting conditions, leading to a total of 1,023 combinations. For each combination, we prepare a human performance of 153 different expressions and render 7 different views for each expression using a customized PBR shader with Blender Cycles. This leads to 1,071 images per combination and over 1 million images in total. 

We optimize each combination to a \Gname asset. For each \Gname asset, the Gaussian points are uniformly initialized on the 4K UV map, with a density of 10 pixels per point for the \textit{Foreface} and 16 pixels per point for the remaining. This leads to a total of 228,083 Gaussian points per asset. Each \Gname asset is optimized with the same hyperparameters and losses as~\citet{kerbl20233d}, with the learning rate of the dynamic shadow vector $\boldsymbol{l}$ set to $1/B$ of the learning rate of the SH base color component. We optimize each \Gname for 30,000 iterations, with around 30 minutes on a NVIDIA RTX 3090.

\section{\name: Instant Gaussian Asset Generation}
\label{sec:gs_gen}
\begin{figure*}[t]
    \centering
    \includegraphics[width=\textwidth]{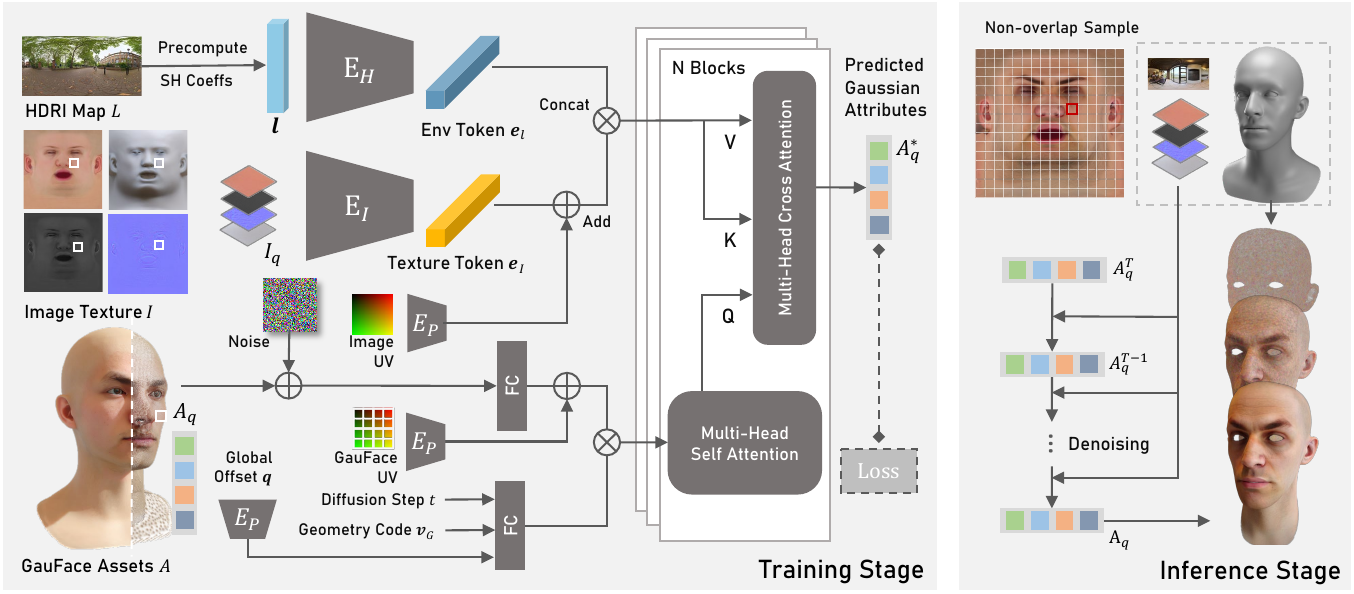}
    \caption{\textbf{TransGS architecture.} We condition \name on the image textures $I$, geometry code $\boldsymbol{v}_G$ and HDRI map $L$, to generate the \Gname asset $A$ in a patch-based manner. \textit{Left:} during training, a random global offset $\boldsymbol{q}$ is sampled, and the corresponding Image patch $I_q$ and \Gname patch $A_q$ are fed to the diffusion transformer. \textit{Right:} at inference, the full \Gname asset can be synthesized in a single pass. }
    \label{fig:gsg_arch}
\end{figure*}

On top of the \Gname representation and dataset, we propose \name, a Gaussian Splatting Translator that maps PBR facial asset to its \Gname counterpart in \textit{seconds}, with close to off-line rendering quality. \name supports inference-level relighting via HDR environment maps and editing by geometry and textures.

As shown in Fig.~\ref{fig:gsg_arch}, conditioning on the image textures $I$, geometry $G$ and lighting information $L$,  our generator $\mathcal{G}$ synthesizes the corresponding \Gname asset $A$:
\begin{equation} 
    A = \mathcal{G}(I, G, L).
\end{equation}

We adopt a diffusion transformer as our backbone and design a patch-based learning strategy to handle the vast number of Gaussian points per asset~(Sec.~\ref{subsec:pdt}). We detail the model condition on PBR facial assets in Sec.~\ref{subsec:gsg_cond} and describe a novel positional encoding to focus the transformer on Gaussian-texture relations in Sec.~\ref{subsec:uvpe}.

\subsection{Patch-based Diffusion Transformer}
\label{subsec:pdt}
\begin{figure}[t]
    \centering
    \includegraphics[width=\linewidth]{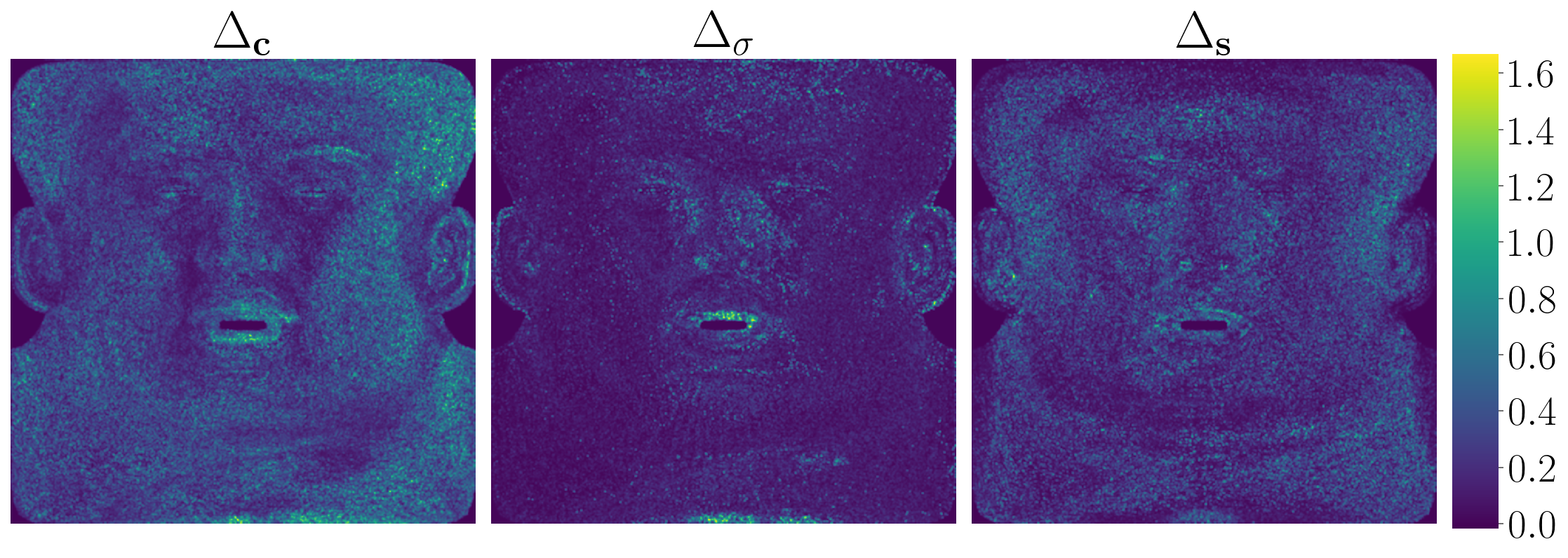}
    \caption{Difference between attributes of two \Gname optimized from the same training images, visualized by plotting attributes to the UV sampling position of Gaussian points. $\Delta_{\boldsymbol{c}}, \Delta_\delta, \Delta_{\boldsymbol{s}}$ are the difference of SH base color, opacity, and specular intensity under two different runs, respectively.} 
    \label{fig:gsg_rand}
\end{figure}

Though we have constrained \Gname assets via multiple design choices~(Sec.~\ref{subsec:pixel}), the mapping between a PBR facial asset under specific lighting, and its \Gname translation is still one-to-many. Fig.~\ref{fig:gsg_rand} visualize the difference between two \Gname assets optimized with the same training data and settings. Gaussian points of two optimized \Gname assets on the same sampled UV position have different attribute values. Thus, we utilize a diffusion process to learn the one-to-many mapping better. Besides, Gaussian assets are essentially structureless point clouds, of which Transformer is preferred for learning correlations \cite{nichol2022point} as it is agnostic to the input sequence order without positional encoding \cite{vaswani2017attention}.

Since each \Gname asset contains a huge number of Gaussian points (roughly 230k), it is impossible to treat all Gaussian points together as a single sequence. Besides, we observe that \Gname assets share similarities in the same UV region across different identities. Thus, we split the UV space into small patches and couple all Gaussian points inside a patch together to serve as the input.  This patch-based processing strategy not only reduces the sequence length of the transformer but also augments the training data by treating small patches independently. 

Specifically, given a Gaussian asset $A$, we sample a random \textit{global offset} $\boldsymbol{q}\in[0, 1]^2$ on the UV plane and collect all the Gaussian points inside the square patch $[\boldsymbol{q}, \boldsymbol{q}+\boldsymbol{P}]$, denoting as a \textit{Gaussian patch} $A_q$. $\boldsymbol{P}$ is a hyperparameter that determines the size of the patch. The denoising process is defined as follows:
\begin{equation}
     A_q = \mathcal{M}(A_q^{t}, t, Y),
\end{equation}
where $A_q^{t}$ is the noised attributes of the Gaussian patch, $t\in[1, ..., T]$ is the diffusion timestep, $Y$ is the condition and $\mathcal{M}$ is the denoising network. $A_q^t$ is mapped to Gaussian tokens $\boldsymbol{e}_A$ through a linear layer, $t$ is mapped to a timestep token $\boldsymbol{e}_T$ similar to DDPM~\cite{ho2020denoising}.

\begin{figure*}[t]
    \centering
    \includegraphics[width=0.98\linewidth]{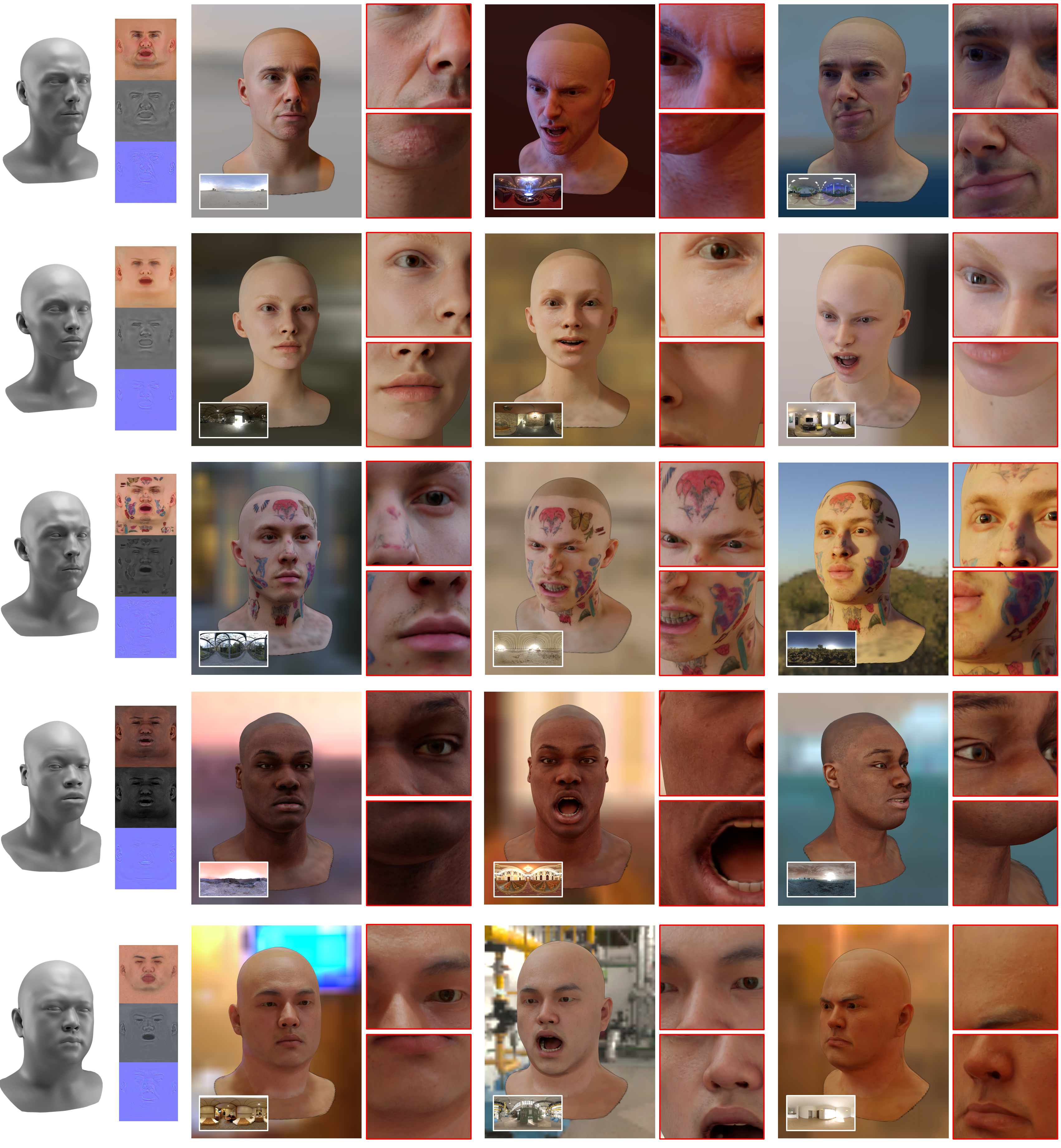}
    \caption{ \textbf{\Gname assets synthesized from \name.} Conditioning on the PBR facial asset, \name generates the \Gname counterpart in \textbf{5 seconds}, with close to off-line rendering quality and supports real-time facial animation and interaction with 30fps@1440p on a \textit{Snapdragon\textsuperscript{\textregistered} 8 Gen 2} mobile phone. The PBR facial assets are obtained from diverse sources.  1-3: Generated from \textit{DreamFace}~\cite{zhang2023dreamface}. 4: Downloaded from web. 5: Scanned from a Light Stage~\cite{debevec2000acquiring}. All figures are rendered under our cross-platform Unity3D \Gname render engine.} 
    \label{fig:gallery}
\end{figure*}

\subsection{PBR Facial Asset Conditioning}
\label{subsec:gsg_cond}
The condition of our transformer, $Y$, includes image textures, geometry, and lighting. 
\textit{Image Textures} are the main conditions of the model. For each PBR facial asset, we concatenate the diffuse, normal, and specular map along the feature dimension to serve as the image condition $I$. For each Gaussian patch $A_q$, we select the \textit{image patch} $I_q$ with the same global offset and UV patch size and map it to image tokens via the image encoder $E_I$, i.e.,  $\boldsymbol{e}_I = E_I(I_q)$.

\textit{Geometry} information is injected via a PCA-based \textit{geometry code}. We utilize a geometry PCA space similar to~\cite {li2020learning}. For each input geometry $G$, the geometry code $\boldsymbol{v}_G$ is obtained by projecting $G$ to the PCA bases. $\boldsymbol{v}_G$ is mapped to the model latent space through a linear layer and then added to the timestep token $\boldsymbol{e}_T$.

\textit{Lighting} is extracted from the HDR environment map by Spherical Harmonics (SH) decomposition~\cite{ramamoorthi2001efficient}. Given an environment map $L$, we compute the 12-order SH coefficients $\boldsymbol{l}$, and map it to a lighting token $\boldsymbol{e}_L$ via a linear layer.  Additionally, to provide detailed shadow information, we bake a low-resolution shadow map $I_s$ and concatenate it with the image condition $I$, to serve as an additional input to the image encoder. 

\subsection{UV Positional Encoding}
\label{subsec:uvpe}
 
We design a novel positional encoding (PE) based on the UV location of inputs to guide the transformer’s attention
toward textures proximal to the Gaussian sampling positions. Specifically, given a UV sampling position  $\boldsymbol{\mu}=(u, v)$, the positional encoding (PE) is calculated as:
\begin{equation} 
\begin{split}
PE(\boldsymbol{\mu}) = \left\{
  \begin{aligned}
    sin(2^{j}\pi u),~&j=4k &cos(2^{j}\pi u),~&j=4k+1 \\
    sin(2^{j}\pi v),~&j=4k+2 &sin(2^{j}\pi v),~&j=4k+3 \\
\end{aligned}  
\right.
\end{split}\quad,
\end{equation}
where $j$ is the dimension of the encoding. The positional encoding is further processed through a projection MLP $E_P$.

Given a pair of patch-based data, we apply positional encoding separately to the Gaussian patch $A_q$, the image patch $I_q$, and the global offset $\boldsymbol{q}$. For each Gaussian patch with a collection of UV sampling positions $\boldsymbol{\mu}_{A, q}$, we add the \textit{relative positional encoding} calculated by $E_P(PE(\boldsymbol{\mu}_{a, q} - \boldsymbol{q}))$ to the Gaussian tokens $\boldsymbol{e}_A$. In addition, given the UV pixel coordinates $\boldsymbol{\mu}_{I, q}$ of the image patch, a similar positional encoding, $E_P(PE(\boldsymbol{\mu}_{I, q} - \boldsymbol{q}))$ is added to the image tokens $\boldsymbol{e}_I$. To inject the global offset information to the denoising network, we add the \textit{global positional encoding}, $E_P(PE(\boldsymbol{q}))$ to the diffusion timestep token $\boldsymbol{e}_T$. 

\begin{figure*}[t]
    \centering
    \includegraphics[width=\linewidth]{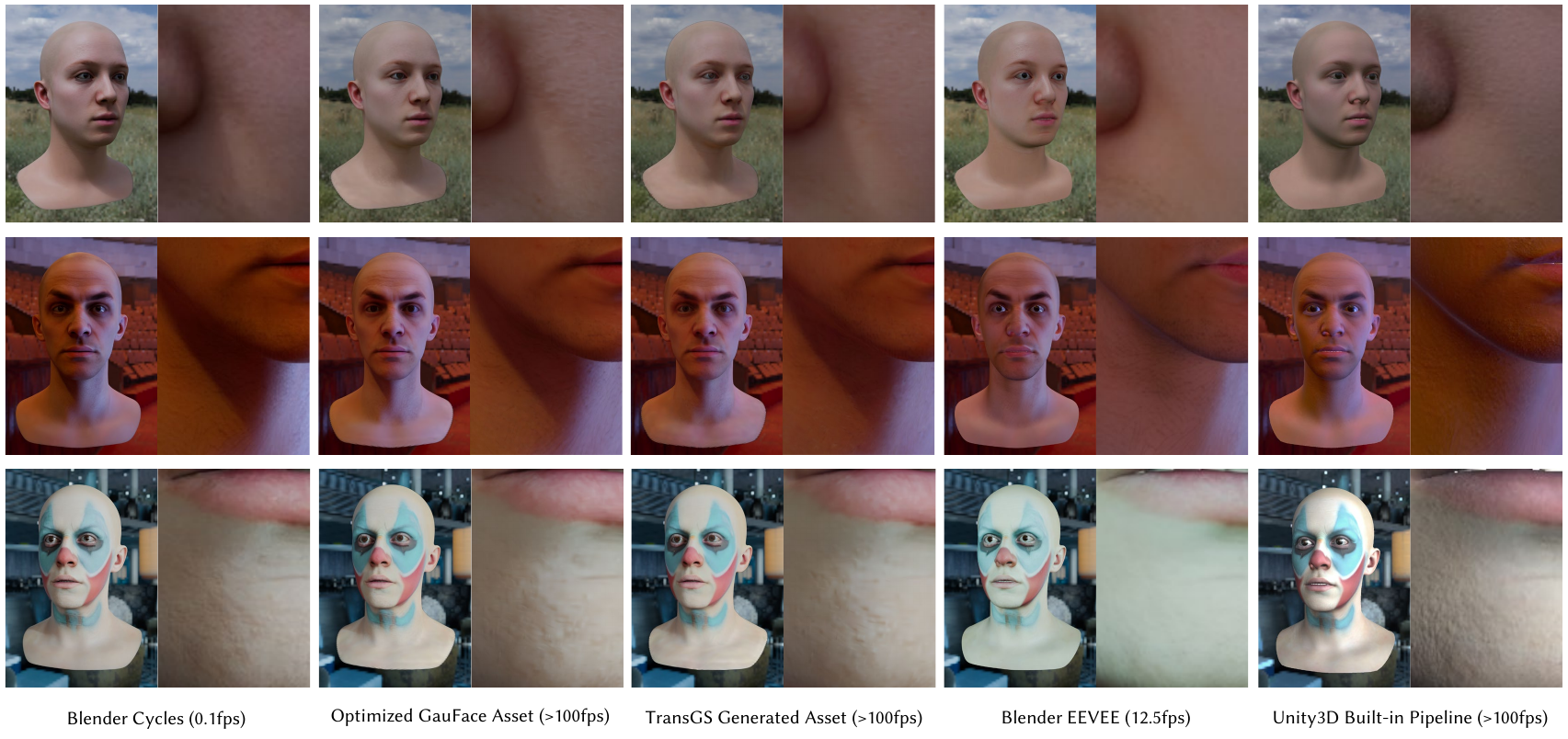}
    \caption{\textbf{Quality comparisons of different rendering pipelines.} We provide the same geometry, image textures, and HDR environment maps to \name and traditional rendering pipelines. All images are rendered with 1080p resolution on a PC with Intel i9-12900K and NVIDIA RTX 4080. Note that the rendering quality of \Gname and \name assets are upper-bounded by Blender Cycles, which provides the ground-truth training images.}
    \label{fig:compare_engine}
\end{figure*}

\subsection{Training}

Our denoising network $\mathcal{M}$ is a decoder-only transformer. During training, we draw random tuples $(A_q, t,  I_q, G, L)$ from the training set and add random Gaussian noise to corrupt $A_q$ to $A_q^t$. After obtaining all input tokens, i.e. Gaussian tokens $\boldsymbol{e}_A$,  Image tokens $\boldsymbol{e}_I$, timestep token $\boldsymbol{e}_T$, and light token $\boldsymbol{e}_L$, we concatenate $\boldsymbol{e}_A$ with $\boldsymbol{e}_T$ to serve as the query to $\mathcal{M}$, and concatenate $\boldsymbol{e}_I$ with $\boldsymbol{e}_L$ to be the key and value. 

We divide the training into two stages: a main stage and a fine-tuning stage. During the main stage, we only consider the simple loss :
\begin{equation}
    \mathcal{L}_s = ||A_q - \mathcal{M}(A_q^{t}, t, I_q, G, L)||_2^2.
\end{equation}

During the fine-tuning stage, we additionally compute the image loss based on the rendered images of the estimated and ground-truth Gaussian assets:
\begin{equation}
    \mathcal{L}_{i} = ||\mathcal{R}(A_q^{*}) - \mathcal{R}(A_q)||_1 + \text{SSIM}\left(\mathcal{R}(A_q^{*}) - \mathcal{R}(A_q)\right).
\end{equation}
where $A_q^{*}$ is the model prediction and $\mathcal{R}$ is the \Gname rendering process. The total loss at this stage is a weighted summation:
\begin{equation}
    \mathcal{L}_\text{total} = \mathcal{L}_s + \lambda \mathcal{L}_i.
\end{equation}

\section{Experiments}
\label{sec:eval}
In this section, we present the experimental results of \name. We first introduce the implementation details, then present a gallery of \name generated assets. Then, we compare the rendering quality of our generated assets against typical offline and online render engines through visualization and a user study. We also compare with neural rendering methods under a time-constrained setting. Finally, we conduct a comprehensive evaluation and ablation studies of our modules and key design choices.

\paragraph{Implementation Details.}
\label{subsec:implementation}
We train our model using the \Gname dataset (Sec.~\ref{subsec:gauface_dataset}). We split it into a training set with 983 assets and a test set with 40 assets. We train separate models for the five facial parts (foreface, backhead, teeth, and left/right eyes, as described in section \ref{subsec:gauface_dataset}) and set the patch size as $\mathbf{P} = (\frac{1}{256}, \frac{1}{256})$, leading to 256 non-overlapping patches per facial component. During training, we randomly sample positions on the UV map as starting points. At inference, we use a consistent set of 256 non-overlapping starting points to cover the entire UV space. 

Our \textit{Foreface} model contains a 12-layer transformer decoder with 512 latent dimensions and a CNN-based image encoder similar to the ControlNet Conditional Embedding~\cite{zhang2023adding}. Models for other facial parts share the same architecture but are scaled down in size. During training, we use 100 diffusion steps and reduce this to 10 steps for inference. We first train the models for 800 epochs in the main stage and finetune it for another 100 epochs with the image loss activated. The training of the \textit{Foreface} model took 5 days on 8 NVIDIA RTX 4090 and others took 1 day each on a single NVIDIA RTX 4090. All modules are trained from scratch in an end-to-end manner. During inference, all patches can be synthesized in a single forward pass, taking 5 seconds in-total on a NVIDIA RTX 4090.

\begin{figure*}[t]
    \centering
    \includegraphics[width=\linewidth]{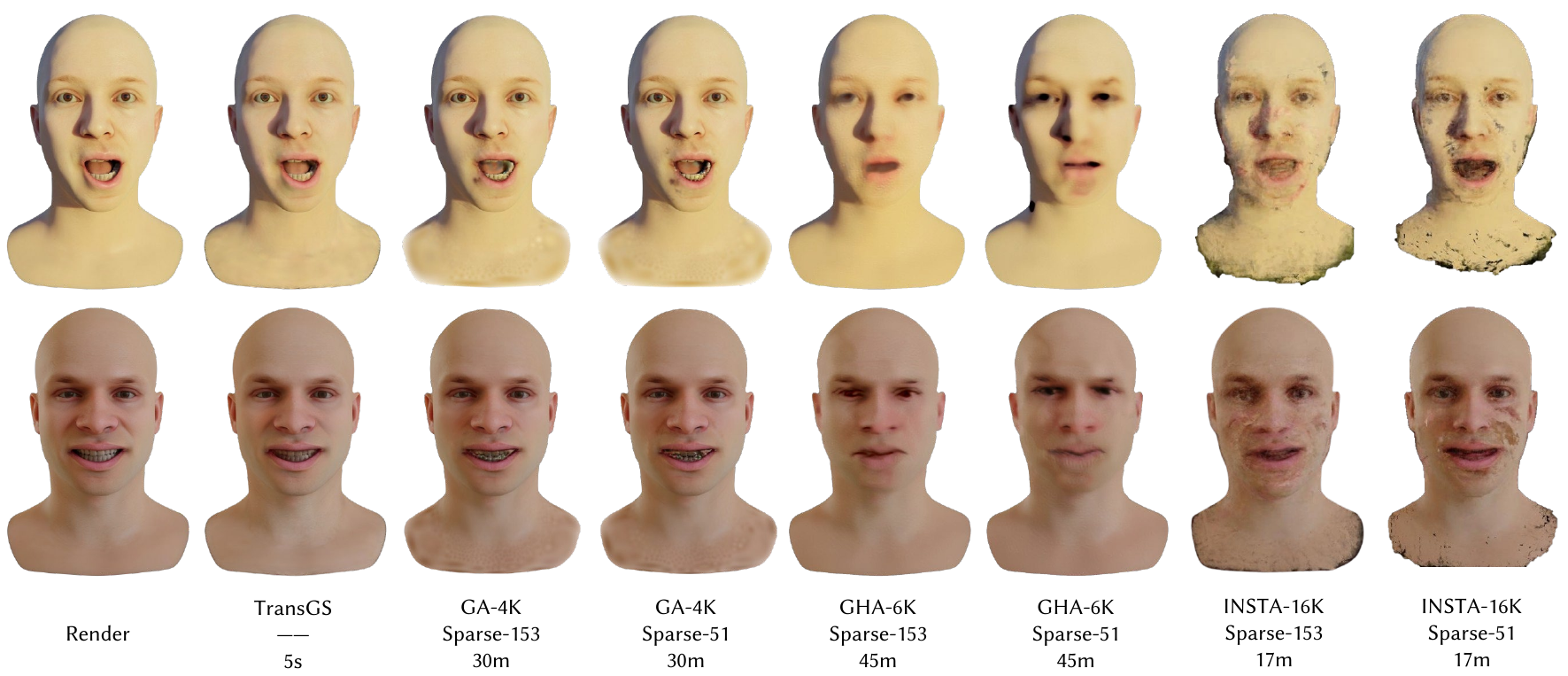}
    \caption{\textbf{Render quality and generation time comparison against other methods}. While \name-generated assets deliver aesthetic rendering quality in 5s, GA fails to model the mouth region under mixed blendshape activations. GHA  and INSTA fail to provide reasonable results due to both data and optimization constraints.} 
    \label{fig:gs_compare_face}
\end{figure*}

Once trained, \name can convert PBR facial assets to \Gname translations in seconds. We showcase the synthesized \Gname assets under Gaussian rendering in Fig.~\ref{fig:gallery}. The \Gname assets are conditionally generated under a variety of lighting conditions and diverse backgrounds, demonstrating the robustness and versatility of our approach. Whether under soft, ambient lighting or stark, directional illumination, the avatars retain their high level of realism, showcasing dynamic shading and realistic light interaction. From neutral, calm expressions to more animated and intense poses, the rendered avatars capture and convey emotions with impressive accuracy.

\subsection{Comparision}

Here, we provide a thorough evaluation of the rendering quality of \name-generated assets. 

\subsubsection{Comparison with CG renders}

 We qualitatively compare the visual quality and the rendering speed of \name synthesized assets against Blender Cycles, Blender EEVEE, and Unity3D. Blender Cycles is an offline ray-tracing rendering engine for production-level visual quality. It also defines the quality upper bound of our method as our dataset relies on Blender Cycles. Blender EEVEE and Unity3D are rasterization-based engines for interactive speed. As shown in Fig.~\ref{fig:compare_engine}, the \Gname assets synthesized by \name deliver aesthetic visual quality comparable to Blender Cycles with natural skin shading from HDR lighting and subsurface scattering, detailed skin textures, sharp shadows, and self-occlusions. In addition, due to the efficient representation, \Gname assets support real-time animation and interaction even on mobile platforms (Fig.~\ref{fig:app_cross}). Note that the visual quality of \Gname assets is upper-bounded by that of Blender Cycles, of which a customized PBR shader is applied. A better shader can potentially lead to better \Gname rendering. 

\paragraph{User study} In the CG industry, rendering quality has always been a subjective measure evaluated by humans. Therefore, we conducted a user study to comprehensively evaluate the rendering quality of \name generated \Gname assets. We designed pairwise comparisons with Blender Cycles, Blender EEVEE, and Unity3D's build-in pipeline, asking the question, "Which image has better facial rendering?" For fairness and comprehensiveness, we included 11 different identities and lighting scenarios. Each questionnaire presented 11 randomly chosen pairs of our renderings alongside one comparison object, with the display order randomized between our renderings and the comparison. 

\begin{figure}[t]
    \centering
    \includegraphics[width=\linewidth]{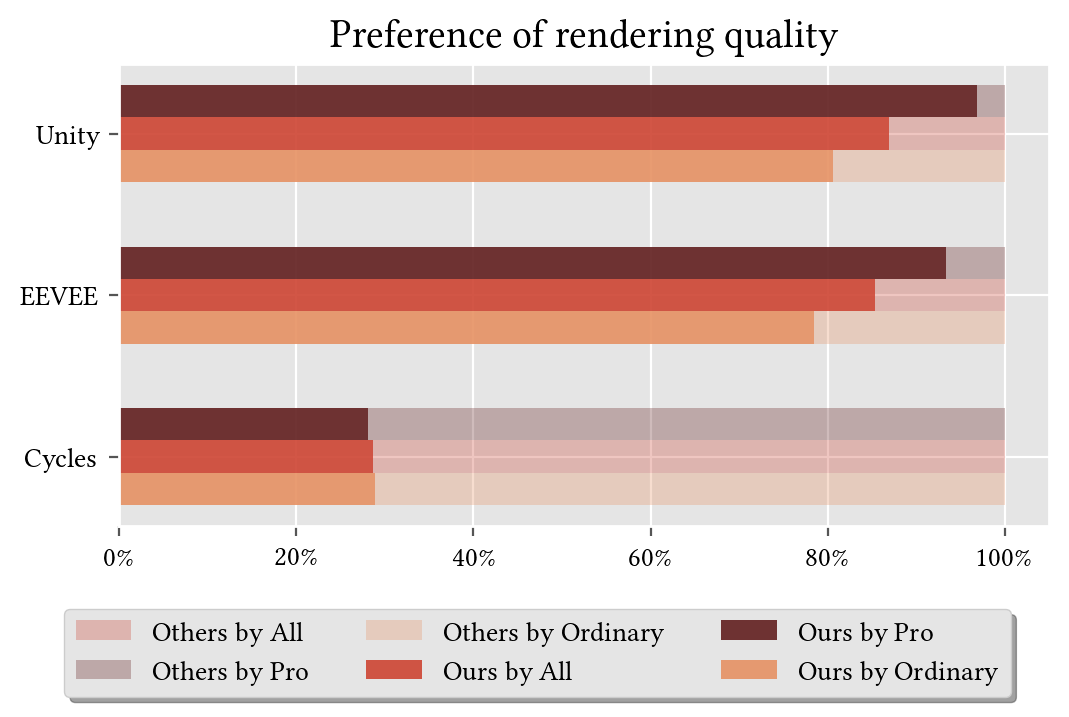}
    \caption{\textbf{Quantitative Results of User Study}. Both professional artists (Pro) and ordinary people (Ordinary) prefer the rendering quality of our \name generated \Gname asset to Unity3D's build-in pipeline (Unity) and Blender EEVEE.}
    \label{fig:user_study}
\end{figure}

We distributed the questionnaire to both professional artists and ordinary people. We received a total of 23 responses from artists and 35 responses from non-artists. As shown in Fig.~\ref{fig:user_study}, over 80\% of the participants preferred our rendering quality to those of Unity3D or Blender EEVEE. As a sanity check, over 70\% participants prefer Blender Cycles to ours, as our rendering quality is upper-bounded by Blender Cycles. It is worth mentioning that professional artists tend to prefer our results over those from Unity and EEVEE, compared to ordinary individuals.

\begin{table}\footnotesize
 \caption{
 \label{tab:gs_compare_func}
 \textbf{Comparison of properties of Gaussian-based facial avatars.} \textit{D} shorts for 'Deformation'. \textit{E} stands for explicit representation and \textit{I} stands for implicit representations. }   
\begin{tabular}{ cccccc } 
 \toprule
 \multirow{2}{*}{Method} & Geometry & \multicolumn{2}{c}{Shading Represenation} &  Relighting  & Instant    \\
 & Representation & \textit{D}-Agnostic & \textit{D}-Dependent & Support & Generation \\
 \midrule

 SA & \textit{E} & \textit{E} & \XSolidBrush & \XSolidBrush & \XSolidBrush \\
 GA & \textit{E} & \textit{E}  & \XSolidBrush & \XSolidBrush & \XSolidBrush \\
 
FA & \textit{I} & \textit{E}  & \XSolidBrush  & \XSolidBrush & \XSolidBrush \\
GBS & \textit{E} & \textit{E}  & \textit{E} & \XSolidBrush  & \XSolidBrush \\
 GHA & \textit{I} & \textit{I} & \textit{I} & \XSolidBrush   & \XSolidBrush\\
 RCA & \textit{I} & \textit{I}  & \textit{I} & \Checkmark  & \XSolidBrush \\
\name & \textit{E} & \textit{E}  & \textit{E} & \Checkmark  & \Checkmark \\
\bottomrule
\end{tabular}

 \end{table}  
 
\subsubsection{Comparison with neural methods}
\label{sec:eval_transgs}
We first compare the properties of \name against other 3DGS-based facial avatars. As illustrated in Tab.~\ref{tab:gs_compare_func}, explicit methods like SplattingAvatar (\textbf{SA})~\cite{shao2024splattingavatar}, GaussianAvatars (\textbf{GA}))~\cite{qian2023gaussianavatars} and 3D Gaussian Blenshapes (\textbf{GBS})~\cite{ma20243d} either omit the Deformation-Dependent Shading effects, or mix that with the Deformation-Agnostic parts (\textbf{GBS}). Implicit methods like FlashAvatar (\textbf{FA})~\cite{xiang2024flashavatar}, Gaussian Head Avatar (\textbf{GHA})~\cite{xu2023gaussian} and Relightable Codec Avatar (\textbf{RCA})~\cite{saito2023relightable} leverage neural networks to enhance representation capacity, which sacrifice 3DGS's affinity with traditional rendering pipelines. Note that all methods except us require per-avatar optimization, which typically consumes minutes or hours to finish on a high-end PC.

To further evaluate the performance and throughput of our novel pipeline, we compare our synthesized results against state-of-the-art optimization-based volume rendering pipelines. Since our method does not require additional data preparation and optimization time, for a fair comparison, we record and limit the runtime of other optimization-based methods.

Specifically, we pick 6 identities from our \Gname test set and design two evaluation variants: Sparse-153 and Sparse-51, which contain 3 and 1 rendered images of individual blendshape activations under random camera positions, respectively. Each case has a test set of another 153 images under different expressions and camera positions.  Since each image requires roughly 10 seconds to render with an NVIDIA RTX 4080, the training set construction time of Sparse-153 takes 25m30s for each identity, and that of Sparse-51 takes 8m30s. 

We compare the rendering quality of our \name synthesized asset against three recent advances in volume rendering pipelines designed explicitly for facial rendering: \textbf{GA}, \textbf{GHA} and \textbf{INSTA}~\cite{zielonka2023instant}.  GA and GHA are Gaussian-based methods, while INSTA is a NeRF variant.  For a fair comparison, we replace GA's FLAME-based representation with our ground-truth geometry to improve its results. GHA and INSTA implicitly relate the volume representation to the geometry via neural networks, so we retain their original pipelines. For GHA, we use images and camera poses from our dataset, optimize the geometry with GHA's network, and perform the two-stage training in their official repo to refine the Gaussians. INSTA requires a segment of monocular RGB video as input; we extract head parameters from the dataset at various angles and expressions, following INSTA's method to obtain facial segmentation and landmarks. We train each method with two versions, varying the number of optimization iterations, denoted with a tail mark (e.g., GA-4K represents the GA method optimized for 4K iterations).

In Table.~\ref{tab:gen_synthetic}, we present the common PSNR, SSIM, and LPIPS metrics, along with the optimization time for each method. Additionally, we compute the lower 90\% quantile of PSNR to emphasize the worst cases. Generally, more optimization iterations result in better performance, with Sparse-153 consistently outperforming Sparse-51, highlighting the data-intensive nature of these methods. However, INSTA is an exception; doubling the optimization iterations leads to worse results due to its facial landmark tracking failing on side or back view renderings. Notably, even the most time-consuming method (GHA-6K on Sparse-153 with 45m) fails to achieve comparable visual quality to the \name synthesized representation.

The qualitative comparisons in Fig.~\ref{fig:gs_compare_face} support the same conclusion. Compared to the Blender Cycles rendered ground-truth, the \name synthesized asset delivers similar visual quality. In contrast, GA struggles with modeling the inner mouth region under mixed blendshape activations due to insufficient training data. GHA exhibits significant blurring effects because its super-resolution module fails to converge within the given time constraints. INSTA fails to accurately model face geometry, resulting in prominent artifacts even in frontal renderings, primarily due to issues with facial landmark tracking.

\begin{table}\footnotesize
\centering
 \caption{
 \label{tab:gen_synthetic}
 \textbf{Visual quality and time consumption of various facial volume rendering methods}. GA, GHA, and INSTA take minutes of data preparation and optimization to reconstruct a scene, while \name generates the \Gname representation with the best rendering quality and takes only 5 seconds on a single NVIDIA RTX 4090. }  
\begin{tabular}{ clccccc } 
 \toprule
 Data & Methods & $\text{PSNR}^\uparrow$ & $\text{SSIM}^\uparrow$ & $\text{LPIPS}^\downarrow$  & $\text{PSNR} (90\%)^\uparrow$   & $\text{Time}^\downarrow$  \\
 \midrule
 
 Sparse-153 
 &GA-4K & 28.02 &  0.9706 & 0.172& 23.87 & 30m \\
 (25m30s)&GA-2K &  25.75 &  0.9666 & 0.181& 21.11& 14m \\
 & GHA-6K & 30.50& 0.9722& 0.112& 27.70&45m \\
 & GHA-3K & 29.50& 0.9703& 0.120& 26.60&24m \\
 & INSTA-16K & 18.31& 0.8080& 0.259& 13.61&17m  \\
 & INSTA-8K & 18.39& 0.8138& 0.255& 13.62&9m  \\
 \hline
 Sparse-51
 &GA-4K & 27.94 & 0.9695 & 0.181& 23.56 &30m  \\
 (8m30s)&GA-2K & 25.99& 0.9667 & 0.172& 21.81&14m  \\

 & GHA-6K & 28.64& 0.9691& 0.112& 25.61&45m \\
 & GHA-3K & 28.03& 0.9677& 0.120& 25.06&24m \\
 & INSTA-16K & 17.62& 0.7945& 0.265& 13.33&17m  \\
 & INSTA-8K & 17.85& 0.8004&  0.261& 13.35&9m  \\
 \hline
 - & \name  & \textbf{35.50} & \textbf{0.9936} & \textbf{0.045}&   \textbf{34.01} & \textbf{5s}   \\
 \bottomrule
\end{tabular}

 \end{table}

\subsection{GauFace Evaluations}
\label{sec:eval_gauface}
In this section, we evaluate the performance of our \Gname representation and its key design choices.

\subsubsection{\Gname representation}

Our \Gname representation serves as a bridge, connecting PBR facial assets to Gaussian representation, and Gaussian representation to generative modeling. Here, we evaluate the performance of \Gname, illustrating its superior rendering quality while keeping a compact and efficient pipeline. 

We use the same six PBR facial assets as described in Section \ref{sec:eval_transgs}, rendering eight images under 153 different expressions and frontal perspectives. For each expression, six images are used for training and two for evaluation. We omit rendering the \textit{Backhead} due to its minimal appearance variation and limited visibility in frontal view renderings. We compare the render quality of \Gname optimized assets against GA, GHA, and INSTA, training all methods for the recommended number of iterations to maximize their representation capabilities.

As shown in Table~\ref{tab:face_synthetic}, our \Gname representation achieves the highest visual quality. GA performs similarly well due to its explicit geometry representation. However, GA's omission of deformation-dependent shading effects slightly reduces its representation power compared to ours. GHA exhibits significant deviations in expressions and positions attributed to the limited fitting capability of its geometric network. INSTA, lacking a clear geometric representation, fails to decouple angles and expressions, resulting in poorer performance. These results support our argument that existing volume-rendering methods, particularly those with implicit representation-geometry relationships, are not directly useable for mapping PBR facial assets with explicit geometry to volume representations.

\begin{table}\footnotesize
\centering
 \caption{
 \label{tab:face_synthetic}
 \textbf{Comparison of visual quality on our synthetic data.} 
  }

\begin{tabular}{ lcccc } 
 \toprule
 Methods & $\text{PSNR}^\uparrow$ & $\text{SSIM}^\uparrow$ & $\text{LPIPS}^\downarrow$  & $\text{PSNR} (90\%)^\uparrow$ \\
 \midrule
INSTA & 25.92  & 0.9122  & 0.1195  & 19.89 \\
GHA & 30.90 & 0.9559 &  0.0870 & 26.45  \\
GA & 45.41 & 0.9912 & 0.0589  & 41.02\\
 
 \Gname  & \textbf{45.67} & \textbf{0.9931} & \textbf{0.0415} & \textbf{41.78}\\
\bottomrule
\end{tabular}

 \end{table}  

\begin{table}\footnotesize
\centering
 \caption{
 \label{tab:face_abl}
 \textbf{Visual quality of \Gname and its ablations.}   
The complex SH Dynamic Shadow Vector (SH DSV) achieves only a 0.3\% improvement in PSNR(90\%) at the expense of a 721.6\% increase in parameters. Without DSV, \Gname cannot deliver high-quality rendering, as evidenced by the PSNR(90\%) dropping below 40. The Normal Delta offers essential flexibility, enabling \Gname to represent visually correct facial deformations accurately.
 }
\begin{tabular}{ lccccc } 
 \toprule
 Methods & $\text{PSNR}^\uparrow$ & $\text{SSIM}^\uparrow$ & $\text{LPIPS}^\downarrow$  & $\text{PSNR} (90\%)^\uparrow$ & $\text{Size} ^\downarrow$ \\
 \midrule
 \Gname  & 45.67 & \textbf{0.9931} & 0.0415 & 41.78 & 106 \\

w/ SH DSV & \textbf{45.72} & \textbf{0.9931} & \textbf{0.0405} & \textbf{41.93} & 871\\ 
w/o DSV & 44.11 & 0.9925 &  0.0428 & 39.68 & \textbf{55} \\
w/o Normal Delta & 39.90 &  0.9893 &  0.0451 & 34.71 & 105 \\
 \bottomrule
\end{tabular}

 \end{table}  
 
\begin{figure}[t]
    \centering
    \includegraphics[width=\linewidth]{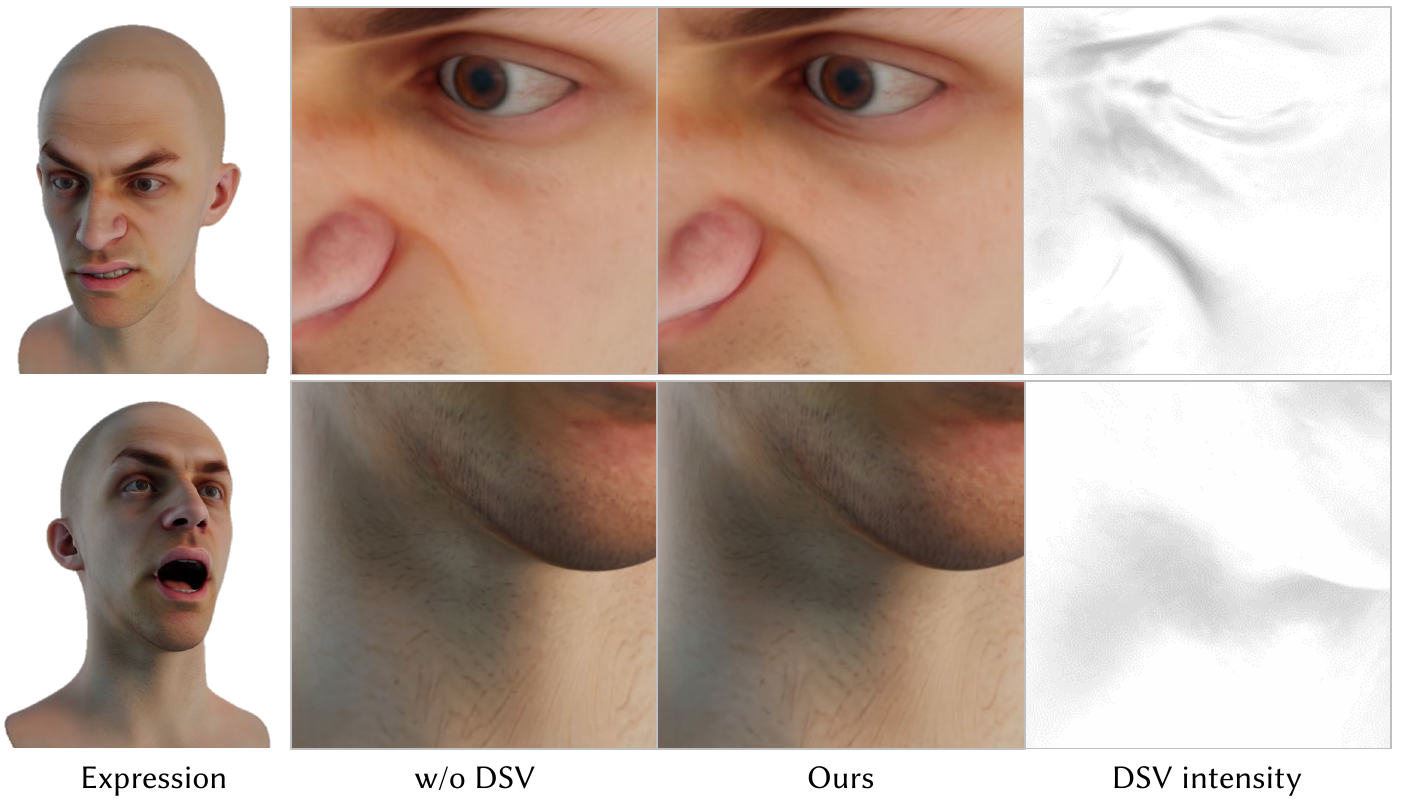}
    \caption{\textbf{Evaluation of Dynamic Shadow Vector } \textit{Top}: the change of local self-occlusions around the nose and eye corners are effectively handled. \textit{Bottom}: It helps to express the change of shadows on the neck caused by mouth-open expression.}
    \label{fig:ab_shadowmap}
\end{figure}

\subsubsection{\Gname ablations} 
\paragraph{Dynamic Shadow Vector} We disentangle the deformation-dependent and deformation-agnostic shading effects via a per-blendshape Dynamic Shadow Vector (DSV) $\boldsymbol{l}$ for each Gaussian Point. In Fig.~\ref{fig:ab_shadowmap}, we qualitatively illustrate the effect of DSV under two different expressions. DSV effectively handles both local self-occlusion changes and shadow variations caused by deformations of distant facial parts.

We further validate the effectiveness of our DSV design by doing an ablation study under the same setting as in Sec.\ref{sec:eval_gauface} without the DSV. Additionally, we provide a more complex DSV variant called SH DSV. Instead of applying the isotropic shadow factor after SH colors computation, SH DSV applies it for every SH component in an anisotropic way. This leads to an additional 765 optimizable parameters for each Gaussian Point. As shown in Tab.~\ref{tab:face_abl}, this anisotropic approach achieves only a 0.3\% performance gain in PSNR~(90\%). However, without DSV, the PSNR~(90\%) drops below 40, indicating a significant degradation in worst-case performance.

\paragraph{Normal Delta} We also evaluate a variant where the Gaussian points are strictly locked to the geometry surface, denoted as 'w/o Normal Delta'. As shown quantitatively in Tab.~\ref{tab:face_abl}, this leads to a significant drop in rendering quality. Gaussian Points need this additional degree of freedom to correctly handle view-dependent and volume effects, including strong directional highlights on the forehand, the shadows inside the mouth and etc.

\begin{table}\footnotesize
\centering
 \caption{
 \label{tab:ablation}
 \textbf{Quantitative results of \name ablations.} Metrics show that the Finetune and UV PE help to increase the quality of \name generation, while PAS plays a fundamental role in the success of generative modeling. }
\begin{tabular}{ lccccc } 
 \toprule
 Methods & $\text{PSNR}^\uparrow$ & $\text{SSIM}^\uparrow$ & $\text{LPIPS}^\downarrow$  & $\text{PSNR} (90\%)^\uparrow$ \\
 \midrule
 Ours  & \textbf{38.69} & \textbf{0.9971} & \textbf{0.033}& \textbf{37.66} & \\
w/o Finetune & 37.23& 0.9962& 0.043& 35.65& \\
w/o UV PE & 35.06& 0.9945&  0.072&  33.45&\\
w/o PAS & 17.24& 0.8416& 0.192& 14.58& \\

 \bottomrule
\end{tabular}

 \end{table}

\subsection{\name ablations}
Here, we evaluate the key design choices related to \name on the \textit{Foreface} model. For each ablation, we compute the PSNR, SSIM, LPIPS, and PSNR~(90\%) metrics of the rendered front images between synthesized \textit{Foreface} Gaussian asset and the test ground-truth \Gname asset, with the same train-test split as described in Sec.~\ref{subsec:implementation}.
\paragraph{Pixel Aligned Sampling} Pixel Aligned Sampling (PAS) provides a strong regularization to the sampling positions of Gaussian points, which is crucial for generative modeling. For ablation, we prepare the same \Gname dataset without applying the PAS, i.e., letting the center of Gaussian points $\boldsymbol{\mu}$ update during optimizations and enable densification and pruning. We train the same \textit{Foreface} model on this PAS disabled dataset. We disable the Positional Encoding since the sampling positions of Gaussian points are not available.

\begin{figure}[t]
    \centering
    \includegraphics[width=\linewidth]{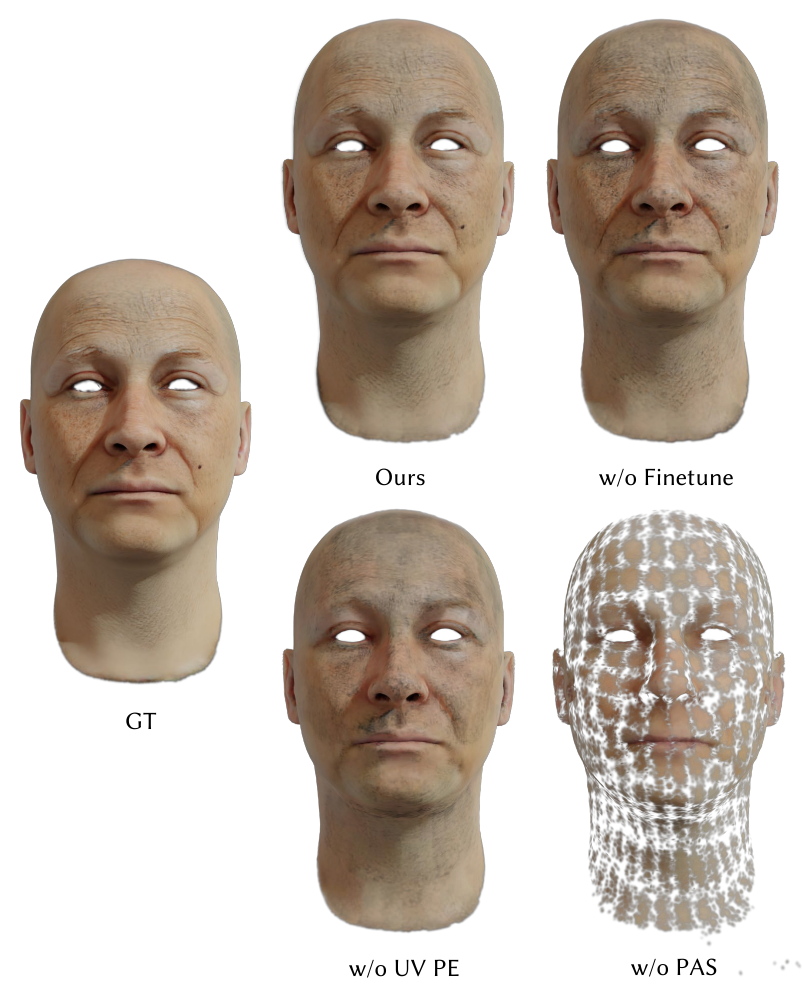}
    \caption{\textbf{Qualitative results of \name ablations.} We compare the frontal rendering of synthesized \textit{Foreface} assets. \name without PAS fails to learn a meaningful distribution of Gaussian points. UV PE helps to capture high-frequency details like melanoma. The Finetune stage provides additional rendering quality. }
    \label{fig:ab_transgs}
\end{figure}

As shown in Table.~\ref{tab:ablation}, without PAS our network fails to model the distribution of Gaussian attributes. The visualization in Fig.~\ref{fig:ab_transgs} illustrates this failure qualitatively. This ablation verifies our hypothesis that the vanilla 3DGS representation is not suitable for generative modeling, and our proposed PAS strategy effectively bridges this gap. 

\paragraph{Positional Encoding} Our UV Positional Encoding (UVPE) helps \name to extract high-frequency details of Image conditions to the generated \Gname asset. Table.~\ref{tab:ablation} indicates an overall degradation of visual quality. As shown in Fig.~\ref{fig:ab_transgs}, the model fails to capture high-frequency details like melanoma without UVPE.

\paragraph{Image Loss Finetuning} In this ablation, after 800 epochs of training in the main stage, we train the model with another 100 epochs while disabling the image loss. As shown in Tab.~\ref{tab:ablation} and Fig.~\ref{fig:ab_transgs}, the fine-tuning stage slightly increases the visual quality of rendered images. 

\section{Applications}
\label{sec:app}

\begin{figure}[h]
    \centering
    \includegraphics[width=\linewidth]{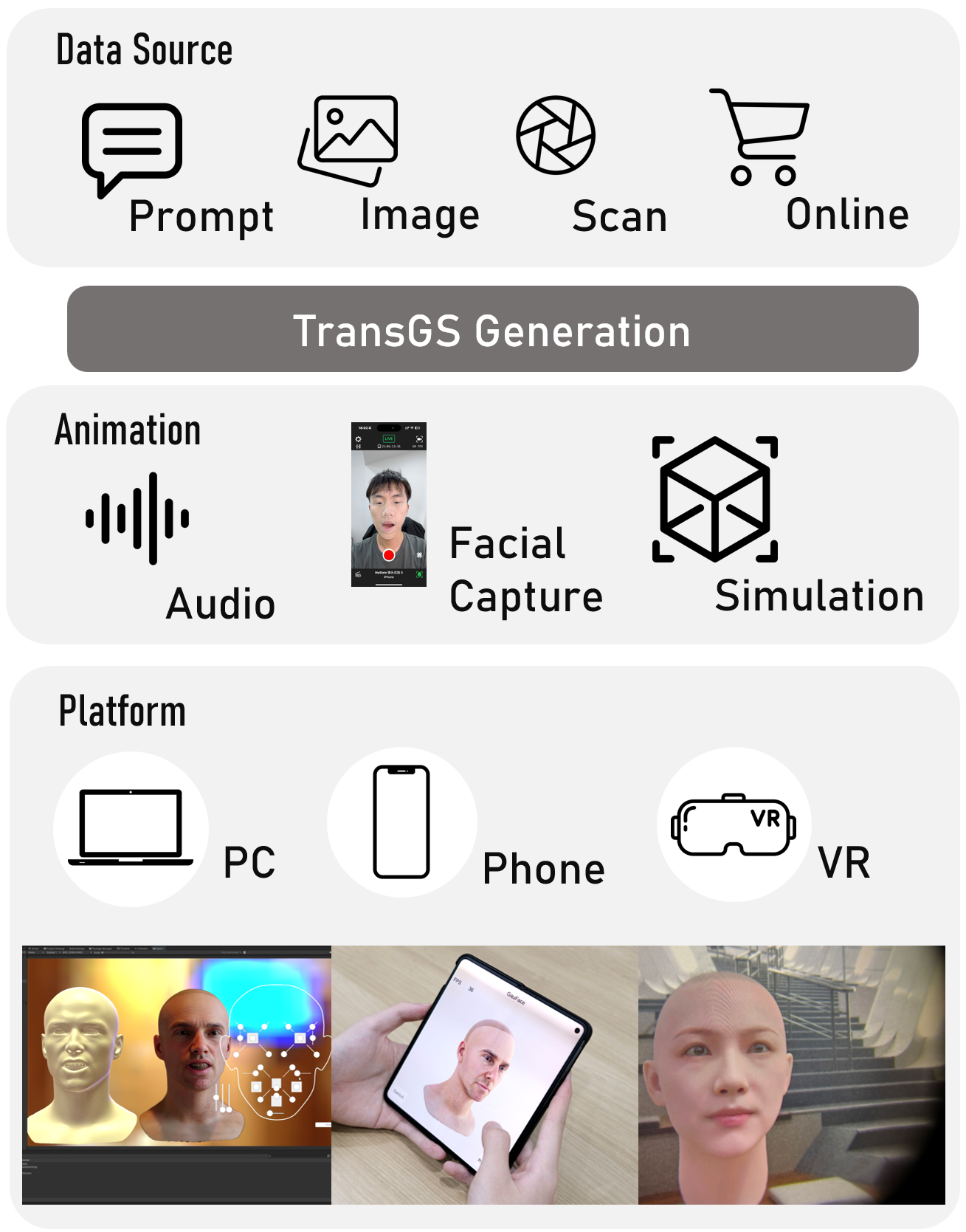}
    \caption{\textbf{Application scenarios.} \textit{Top:} \name accepts PBR facial asset conditions from diverse data sources, including prompt or image generated, scanned,  online downloaded assets. \textit{Middle:} \name generated \Gname asset can be controlled via various inputs, e.g., audio-driven signal, ARKit facial capture, and physical simulation, with cross-platform real-time rendering performance (\textit{Bottom}).}
    \label{fig:app_cross}
\end{figure}

Our \Gname and \name pipeline empower a wide range of applications (Fig.~\ref{fig:app_cross}). On the one hand, \name integrates PBR facial assets from various sources with Gaussian representations. As shown in Fig.~\ref{fig:gallery}, whether the assets are generated, scanned, or downloaded from the internet, \name can transform them into high-quality Gaussian assets under specific lighting conditions. On the other hand, \Gname's explicit association with geometry enables support for various driving forms in traditional CG animation, such as ARKit facial capture, physics simulation, and audio-driven methods. 
Thanks to the explicit and efficient \Gname representation, unprecedented rendering quality with real-time animation and interaction can be delivered to mobile platforms. Please refer to our supplementary video for various application showcases.
\subsection{Cross-Platform Real-time Rendering}

We implement a \Gname renderer using Unity3D's built-in render pipeline. The explicit representation of \Gname allows us to leverage Unity's multi-platform compilation support, enabling high-quality \Gname rendering and animation effects on platforms supporting DirectX 12 and the Vulkan Graphics API, including Windows Standalone, Android, and VR headsets. In a basic scene with a physical camera, a well-designed skybox, and a \Gname object containing 110K Gaussian points, we achieved 500+ fps on an NVIDIA RTX 4080 GPU during animations. On the Android platform, we reached 30 fps on a \textit{Snapdragon\textsuperscript{\textregistered} 8 Gen 2} mobile platform at 1440p resolution. Similar performance is delivered to a Meta Quest Pro VR headset with on-chip computation. Please refer to our supplementary video for the live demo.

\begin{figure}[t]
    \centering
    \includegraphics[width=\linewidth]{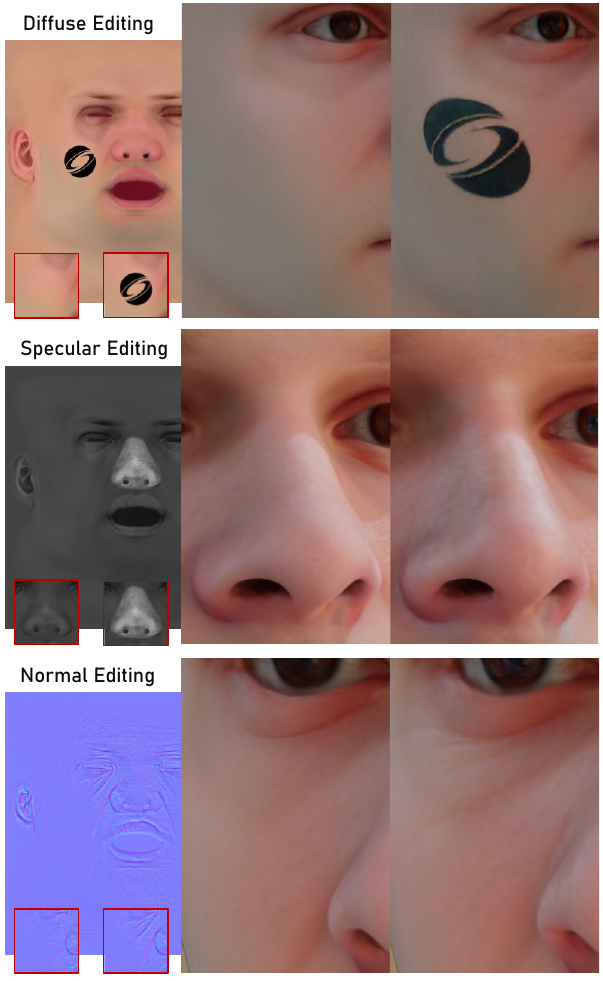}
    \caption{ \textbf{Various patch-wise editing.} From left to right: texture editing; rendering of the original \Gname asset; rendering of the edited \Gname asset. All \Gname assets are synthesized by \name. We can modify the diffuse (top), normal (middle), and specular (bottom) maps, and \name can faithfully transfer these details to the \Gname asset, with natural shading under the conditioned lighting. }
    \label{fig:app_mani}
\end{figure}

\subsection{Interactive \Gname Editing}
Since \name is conditioned on traditional assets, any modifications made to the traditional assets can also be transferred to \Gname through the model. Furthermore, because our generative pipeline is patch-based, the network only needs to handle local modification information, making the entire updating process compact and efficient.

We showcase in Fig.~\ref{fig:app_mani} three editing scenarios. In the first row of the image, we overlaid a previously unseen logo onto the diffuse map. As shown in the rightmost render, the logo has been successfully transferred to the Gaussian asset, maintaining high clarity and sharp edges while seamlessly integrating into the lighting environment. In the second row, we brightened the specular map for the character's nose. Compared to the previous more diffuse rendering (middle), the enhanced nose (right) now exhibits more pronounced highlights. In the third row, we modified the normal map to add new wrinkles around the character's left eye and nose. The modified Gaussian asset is able to render these wrinkles with a high degree of realism and natural detail, showcasing a rich and detailed appearance.

\section{Conclusion}
\label{sec:limits}
We have presented \name, a novel method for translating PBR facial assets into 3D Gaussian Splatting representations instantaneously, supporting
relighting and high-quality real-time interaction across different rendering platforms.  To achieve this, we proposed \Gname, a novel Gaussian representation for efficient rendering of facial interaction. Leveraging strong geometric priors and constrained optimization, \Gname ensures neat and rigid representation, bridging the gap between PBR facial assets and Gaussian representation and between Gaussian representation and generative modeling. We then designed a diffusion transformer that translates PBR facial assets to \name assets. We proposed a novel pixel-aligned sampling scheme and UV positional encoding to ensure the throughput and rendering quality of \name generation.
Through \name, PBR facial assets can be rapidly converted into efficient Gaussian assets, enabling more realistic real-time facial expression rendering under different lighting conditions across various platforms. This offers new possibilities for immersive interactive experiences, enhanced storytelling and etc. Additionally, \name endows Gaussian assets with the capability to be modified and driven by traditional CG pipelines, greatly expanding their application scope.

\paragraph{Limitations.} As an initial attempt, our method has several limitations. For instance, we did not perform Gaussian Splatting modeling and generation for hair, which is a complex task beyond our current scope. Our relighting capability, driven by \name, while fast, cannot be applied in real-time applications. Therefore, exploring ways to enhance our \Gname representation for real-time relighting is a promising direction. Although our approach leverages PBR facial assets and uses Blender Cycles, a physically based rendering engine, to provide ground-truth images, the rendering of our generated \name assets is not physically accurate because we do not explicitly encode physically correct shading or apply any physical constraints to the generator. Additionally, we model different facial subparts separately, neglecting the correlations of shading between components. Lastly, the rendering quality of our \name generated assets is constrained by our customized Blender shader; improving this shader could potentially enhance the visual quality of \name synthesized assets.

\paragraph{Potential ethical implications} \name's ability to create highly realistic facial Gaussian avatars raises concerns about consent and control over one's digital representation. Individuals may not have control over how their digital likeness is used, leading to potential harm if their likeness is used without permission or in ways that are harmful or misleading. Additionally, there may be societal implications related to the uncanny valley effect and its impact on human perception and interaction. Highly realistic facial avatars may blur the lines between virtual and real identities, potentially affecting social dynamics and human relationships. It is important for future work to consider the privacy protection and intellectual property control capabilities of \Gname assets and the personal data required for generating \Gname assets should only be used with explicit permission and authorization to avoid infringement of personal privacy rights.

\section{Acknowledgements}
\label{sec:ack}
This research is supported by Innovation and Technology Commis-
sion (Ref:ITS/319/21FP) and Research Grant Council (Ref: 17210222),
Hong Kong.

\bibliographystyle{ACM-Reference-Format}
\bibliography{ref}

\end{document}